\documentclass[usegraphicx,useAMS,usenatbib]{mn2e}

\usepackage{amssymb}
\usepackage{graphicx} 
\usepackage{epsfig}
\usepackage{amsmath} 
\usepackage{txfonts}

\def\mnras{{Mon.~ Not.~ R.~ Astron.~ Soc.~}}

\def\prd{{Phys.~ Rev.~ D.~}}

\def\apj{{Astrophys.~ J.~}}
\def\apjs{{Astrophys.~ J.~ Suppl.~}}
\def\apjl{{Astrophys.~ J.~ Lett.~}}

\def\aj{Astronomical Journal}

\def\mnras{{MNRAS}}

\def\prd{{PRD}}

\def\apj{{ApJ}}
\def\apjs{{ApJS}}
\def\apjl{{ApJL}}
\def\aap{{A\&A}}

\def\physrep{{Phys.~ Rep.~}}

\newcommand{\be}{\begin{equation}}
\newcommand{\ee}{\end{equation}}
\newcommand{\ba}{\begin{eqnarray}}
\newcommand{\ea}{\end{eqnarray}}

\def\pp1{{\prime}}
\def\pp2{{\prime\prime}}

\def\2D{{\rm 2D}}

\def\1Loop{{\rm 1Loop}}

\def\fun#1#2{\lower3.6pt\vbox{\baselineskip0pt\lineskip.9pt
        \ialign{$\mathsurround=0pt#1\hfill##\hfil$\crcr#2\crcr\sim\crcr}}}

\newcommand{\vn}{\mbox{\bf {n}}}

\newcommand{\vk}{\mbox{\bf {k}}}

\newcommand{\vq}{\mbox{\bf {q}}}

\newcommand{\vnh}{\hat{\mbox{\bf {n}}}}

\newcommand {\lessim} {\ {\raise-.5ex\hbox{$\buildrel<\over\sim$}}\ }

\usepackage{multirow}
         
\def\plotancho#1{\includegraphics[width=18cm]{#1}}
\def\refe@jnl#1{{#1}}
\def\aj{\refe@jnl{Astron.~J.}}                  
\def\araa{\refe@jnl{Annu.~Rev.~Astron.~Astrophys.}}
\def\apj{\refe@jnl{Astrophys.~J.}}                 
\def\apjl{\refe@jnl{Astrophys.~J.~Lett.}}          
\def\apjs{\refe@jnl{Astrophys.~J.~S.~S.}}          
\def\aap{\refe@jnl{Astron.~Astrophys.}}            
\def\mnras{\refe@jnl{Mon.~Not.~R.~Astron.~Soc.}}   
\def\prd{\refe@jnl{Phys.~Rev.~D}}        
\def\fcp{\refe@jnl{Fund.~Cos.~Phys.}}  
\def\physrep{\refe@jnl{Phys.~Rep.}}
\def\physlett{\refe@jnl{Phys.~Lett.}}

\DeclareTextSymbol{\degre}{OT1}{23}

\title[BOSS constraints on the ISW]{The SDSS-III Baryonic Oscillation Spectroscopic Survey: Constraints on the Integrated Sachs Wolfe effect}

\author[{\it Hern\'andez-Monteagudo et al.}]
{\parbox{\textwidth}
       {Carlos Hern\'andez-Monteagudo$^{1,2}$\thanks{chm@cefca.es},
                Ashley J. Ross$^{3}$, Antonio Cuesta$^{4}$, 
                Ricardo G\'enova-Santos$^{5,6}$, 
                Jun-Qing Xia$^{7}$,
                Francisco Prada$^{8,9,10}$, Graziano Rossi$^{11}$, Mark Neyrinck$^{12}$, Matteo Viel$^{13}$, 
                Jose-Alberto Rubi\~no-Martin$^{5,6}$, Claudia G. Sc\'occola$^{5,6}$, 
                Gongbo Zhao$^3$, Donald P. Schneider$^{14}$, Joel R. Brownstein$^{15}$, 
                Daniel Thomas$^{3}$  and Jonathan V. Brinkmann$^{16}$}\vspace{0.4cm}\\
{\parbox{\textwidth}{         
         {$^1$ Centro de Estudios de F{\'\i}sica del Cosmos de Arag\'on (CEFCA), 
           Plaza de San Juan, 1, planta 2, E-44001, Teruel, Spain}\\
         {$^2$ Max-Planck Institut f\"ur Astrophysik, 
           Karl Schwarzschild Str.1, D-85741,
           Germany\\
           $^3$ Institute of Cosmology and Gravitation, Dennis Sciama Building, University of Portsmouth, Portsmouth, PO1 3FX, UK\\
           $^4$ Yale Center for Astronomy and Astrophysics, Yale University, New Haven, CT 06511, USA\\
           $^5$ Instituto de Astrof{\'\i}sica de Canarias (IAC), C/V{\'\i}a
L\'actea, s/n, E-38200, La Laguna, Tenerife, Spain. \\
           $^6$ Dpto. Astrof{\'\i}sica, Universidad de La Laguna (ULL), E-38206
La Laguna, Tenerife, Spain.\\
	$^7$Key Laboratory of Particle Astrophysics, Institute of High Energy Physics, Chinese Academy of Science, P.O. Box 918-3, Beijing, 100049, P.R. China\\
           $^8$ Campus of International Excellence UAM+CSIC, Cantoblanco, E-28049 Madrid, Spain \\
	 $^9$ Instituto de F\'{\i}sica Te\'orica, (UAM/CSIC), Universidad Aut\'onoma de Madrid, Cantoblanco, E-28049 Madrid, Spain \\
	$^{10}$ Instituto de Astrof\'{\i}sica de Andaluc\'{\i}a (CSIC), Glorieta de la Astronom\'{\i}a, E-18008 Granada, Spain\\
	$^{11}$CEA, Centre de Saclay, Irfu/SPP, F-91191 Gif-sur-Yvett, France\\
	$^{12}$Department of Physics and Astronomy, The Johns Hopkins University, Baltimore, MD 21218, USA\\
	$^{13}$ INAF Osservatorio Astronomico, via Tieopolo 11, I-34143 Trieste, Italy\\
	$^{14}$ Department of Astronomy and Astrophysics, The Pennsylvania State University, University Park, PA 16802, USA\\
	$^{15}$Department of Physics and Astronomy, University of Utah, 115 S 1400 E, Salt Lake City, UT 84112, USA\\
	$^{16}$Apache Point Observatory, 2001 Apache Point Road, Sunspot, NM 88349-0059, USA\\
	 }
       }
       }}
       
\begin{document}

\maketitle

\begin{abstract}
 
In the context of the study of the Integrated Sachs Wolfe effect (ISW), we construct a template of the projected density distribution up to redshift $z\simeq 0.7$ by using the Luminous Galaxies (LGs) from the eighth Sloan Digital Sky Survey (SDSS) data release (DR8). We use a photometric redshift catalogue trained with more than a hundred thousand galaxies from the Baryon Oscillation Spectroscopic Survey (BOSS) in the SDSS DR8 imaging area covering nearly one quarter of the sky. We consider two different LG samples whose selection matches that of SDSS-III/BOSS: the low redshift sample (LOWZ, $z\in [0.15,0.5]$) and the constant mass sample (CMASS, $z\in[0.4,0.7]$). 
When building the galaxy angular density templates we use the information from star density,  survey footprint, seeing conditions, sky emission, dust extinction and airmass to explore the impact of these artifacts on each of the two LG samples. In agreement with previous studies, we find that the CMASS sample is particularly sensitive to Galactic stars, which dominate the contribution to the auto-angular power spectrum below $\ell=7$. Other potential systematics affect mostly the very low multipole range ($\ell\in[2,7]$), but leave fluctuations on smaller scales practically unchanged. The resulting angular power spectra in the multipole range $\ell\in[2,100]$ for the LOWZ, CMASS and LOWZ+CMASS samples are compatible with linear $\Lambda$CDM expectations and constant bias values of $b=1.98 \pm 0.11$, $2.08\pm0.14$ and $1.88\pm 0.11$, respectively, with no traces of non-Gaussianity signatures, i.e., $f_{\rm NL}^{\rm local}=59\pm 75$ at 95\,\% confidence level for the full LOWZ+CMASS sample in the multipole range $\ell\in[4,100]$. 
After cross-correlating WMAP-9yr data with the LOWZ+CMASS LG  projected density field,
the ISW signal is detected at the level of 1.62--1.69$\,\sigma$. While this result is in close agreement with theoretical expectations and predictions from realistic Monte Carlo simulations in the concordance $\Lambda$CDM model, it cannot rule out by itself an Einstein-de Sitter scenario, and has a moderately low signal compared to previous studies conducted on subsets of this LG sample. We discuss possible reasons for this apparent discrepancy, and point to uncertainties in the galaxy survey systematics as most likely sources of confusion.
   \end{abstract}

\date{Received XXXX; accepted YYYY}
 
\begin{keywords}
  cosmology: observations -- cosmic microwave background -- large-scale
  structure of the Universe 
\end{keywords}

\section{Introduction}

The growth and evolution of density and potential perturbations in an almost perfectly isotropic and homogeneous universe can be described in General Relativity by the family of Friedmann-Robertson-Walker (FRW) metrics \citep[see, e.g.,][]{LifshitzKhalatnikov1963,Mukhanovetal1992}. Despite their requirements on isotropy and homogeneity, these models leave room for the study of small {\it clumps} or inhomogeneities that eventually give rise to the complex structure of today's universe. Indeed, under the assumption that initial fluctuations in the metric are small, it is possible to write a linear theory of perturbations describing the interplay and evolution of small inhomogeneities in a FRW background \citep[see, e.g.,][and references therein]{MaBertschinger1995}. 

In particular, FRW models show that, on the large scales within the horizon where the Poisson equation holds, the growth rate of the gravitational potentials ($D_{\phi}(t)$) is proportional to the time derivative of the ratio $D_{\rho}(t) / a(t)$, where $a(t)$ is a FRW metric scale factor describing the expansion of scales in the universe, and $D_{\rho}(t)$ is a function describing the (linear) growth of matter density in-homogeneities.   In the special case of an Einstein-de Sitter universe $D_{\rho}(t) = a(t)$ and hence gravitational potentials on large scales should remain constant \citep[see, e.g.,][and references therein]{MaBertschinger1995}. However, deviations from this scenario will make $D_{\rho}(t)\neq a(t)$ and thus variations in the large scale gravitational potentials will arise. Photons of the Cosmic Microwave Background (CMB) crossing those potentials will leave  a {\em different} potential than they entered, and the difference will appear in the form of a gravitational blue/redshift that is known as the late Integrated Sachs-Wolfe effect, \citep[hereafter ISW,][]{SachsWolfe1967}.

If dark energy is present at low redshift, and the Universe is expanding faster than if it were matter-dominated, then linear-scale gravitational potential wells shrink with time, and CMB photons leave potentials that have become shallower during their fly-by.
This results in more energetic CMB photons emerging out of overdensities, and hence the ISW temperature anisotropies of the CMB should be positively correlated to the distribution of gravitational potential wells. \citet{CrittendenTurok1996} first noticed the fact that those wells are probed by halos and galaxies, and that the distribution of these structures can be cross-correlated to the CMB anisotropies in order to distinguish the ISW component from the intrinsic CMB anisotropies generated at the surface of last scattering, at $z\simeq 1,050$.

As soon as the first data sets from {\it Wilkinson Microwave Anisotropy Probe} (hereafter WMAP) were released, several works
claimed detections of the ISW at various levels of significance (ranging from 2$\,\sigma$ to $>4\,\sigma$) after cross-correlating WMAP data
with different galaxy surveys,
\citep[][]{Scrantonetal2003short,Fosalbaetal2003,BoughnCrittenden2004,
  Fosalbaetal2004,Noltaetal2004,Afshordietal2004,Padmanabhanetal2005c,
  Cabreetal2006,Giannantonioetal2006,Vielvaetal2006,McEwenetal2007}.
However, other authors \citep{HernandezMonteagudoetal2006,Rassatetal2007,2009ApJ...701..414G,Bielbyetal2010,LopezCorredoiraetal2010,HernandezMonteagudo2010,PeacockFrancis2010,Sawangwitetal2010,2013MNRAS.431L..28K}, reported lower (and/or lack of) statistical significance and/or presence of point source contamination in the cross-correlation between different galaxy surveys and WMAP CMB maps. In \citet{HernandezMonteagudo2008} it was shown that the cross-correlation of the ISW with galaxy templates should be contained on the largest angular scales ($\ell<60$--$80$), with little dependence on the galaxy redshift distribution, and that most of such signal should arise in the redshift range $z\in [0.2,1.0]$. However, most of the surveys used in ISW--galaxy cross correlation studies were either shallow and/or had more anisotropy power than predicted by the models precisely on the large angular scales sensitive to the ISW. 
  
 The issue of this power excess present at large scales, when computing the auto-correlation function or auto power spectrum in galaxy surveys, has been relevant for some time \citep[see, e.g.,][]{Hoetal2008,HernandezMonteagudo2010,Thomasetal2011,
  Giannantonioetal2012}. The shape of this excess is similar to that expected in models with high levels of local Non-Gaussianity (NG), and thus it has motivated claims of NG detection in the past, \citep[see, e.g.,][]{Xiaetal2010,Xiaetal2010b}. Recently it has been recognized that the presence of systematics in the galaxy survey data may be largely responsible for this extra power \citep{Hutereretal2012}, and thus latest constraints on NG are consequently weakened, \citep[see][]{Rossetal2012}. Unfortunately, the impact that these systematic effects - giving rise to extra power on large scales - may have on the ISW signal has not yet been properly addressed. The last studies on the subject \citep{Giannantonioetal2012,Schiavonetal2012} have pointed out the presence of this excess, but they did not quantified its impact on the expected ISW signal-to-noise ratio, or galaxy cross-correlation analysis, i.e. either if due to NG or to systematics. Since this unexpected high level of anisotropy on the large scales is present in different galaxy surveys, it should also affect those works combining different surveys \citep[e.g.][]{Giannantonioetal2008,Hoetal2008,Giannantonioetal2012}. The parallel analysis of \citet{Giannantonioetal2013} has nearly simultaneously obtained similar results in terms of ISW significance using a similar data set to our own.
    
The Sloan Digital Sky Survey-III \citep[SDSS-III,][]{Eisensteinetal2011} Baryonic Acoustic Oscillation Spectroscopic Survey \citep[BOSS,][]{Dawsonetal2013} has calibrated a number of potential systematics (star density, seeing, sky emission, etc) that have to be used when building templates of galaxy angular number density, \citep{Rossetal2012,Hoetal2012}. In this work we make use of all the information related to potential systematics to produce {\em corrected} templates of the number density of Luminous Galaxies from two different BOSS samples: the low redshift sample (LOWZ, $z\in [0.1,0.5]$) and the constant mass sample, (CMASS, $z\in[0.4,0.7]$). These templates are used as tracers of moderate-redshift gravitational potential wells, in order to be correlated with the CMB maps produced by WMAP. In Section ~\ref{sec:BOSS} we describe the SDSS Data Release 8 (DR8) and SDSS-III/BOSS data used for the LG template construction, and in Section ~\ref{sec:WMAP} the CMB data from WMAP used in the cross-correlation analysis. In Section ~\ref{sec:theorpred} we describe the theoretical predictions for the level of LG -- CMB cross-correlation to be induced by the ISW effect. The statistical methods used in our cross-correlation analysis are described in Section ~\ref{sec:statmethods}, and its results are presented in Section ~\ref{sec:results}. These results are discussed in Section ~\ref{sec:disconcl}, where conclusions are also presented. The detailed explanation of the systematic corrections are provided in Appendix \ref{sec:syst}. These corrections must be taken into account  when computing and interpreting the angular power spectrum of the LG samples, as described in Appendix \ref{sec:aps}. This appendix also includes tests for remaining residuals systematic and an assessment on the compatibility of the data to the Gaussian hypothesis (i.e., limits on the non-Gaussian parameter $f_{\rm NL}^{\rm local}$).

Unless stated otherwise, we employ as a reference a flat $\Lambda$CDM cosmological model
consistent with WMAP9 \citep{Hinshawetal2012}, with
density parameters $\Omega_b=0.04628$, $\Omega_{\rm cdm}=0.24022$,
$\Omega_{\Lambda}=0.7135$, reduced Hubble constant $h=0.6932$, scalar
spectral index $n_{\rm S}=0.9608$, optical depth to the Last Scattering
Surface $\tau_T=0.081$, and rms of relative matter fluctuations in spheres of
$8\,h^{-1}$\,Mpc radius $\sigma_8=0.82$.

\section{BOSS LG sample}
\label{sec:BOSS}

The galaxy sample we use is based on a photometric selection of galaxies from the SDSS Data Release 8 \citep[][ DR8]{Yorketal2000,DR8}. This survey obtained wide-field CCD photometry (\citealt{Gunn98,Gunn06}) in five passbands ({\it u, g, r, i, z}; e.g. \citealt{Fuku96}). The DR8 imaging data covers 14,555 deg$^2$, more than one third of which is in the Southern Galactic cap. This large sky area provides an unprecedented signal-to-noise ratio for ISW studies as compared to previous analysis in the literature, as shown in Section 4.

We select galaxies using the same color and magnitude cuts defined in \citet{Eisensteinetal2011}, Eqs. 1-5. This includes a selection for a lower redshift ($z \lessim 0.4$) sample denoted `LOWZ' and higher redshift ($0.4 \lessim z \lessim 0.7$) composed of galaxies that are approximately stellar mass limited and denoted 'CMASS'. These photometric selections yield a total of 2.4 million galaxies (after removing any duplicate galaxies that appear in both selections), one third of which are in the LOWZ sample. Photometric redshifts, galaxy/star probabilities, and recommendations for measuring the angular clustering for the CMASS sample were described by \cite{Rossetal2011} and the angular clustering of the CMASS sample was studied further in \cite{Hoetal2012}. For the LOWZ sample we also generate photometric redshifts \citep{Boltonetal2012}, as we did for CMASS in \cite{Rossetal2011}, using BOSS spectroscopic redshifts as a training sample \citep{Smeeetal2012}. These photometric redshifts are trained using the neural network code ANNz \citep{Firth03}. We remove LGs with $z_{phot} < 0.15$, as such galaxies are expected to have minimal contribution to the ISW signal while they can be more easily confused with other types of galaxies \citep{Parejkoetal2013}.  This cut removes $1.2\times 10^5$ galaxies. \footnote{The resulting catalogues can be found at {\tt ftp://ftp.cefca.es/people/chm/boss\_lgs/}}

The SDSS DR8 imaging area deemed suitable for galaxy clustering studies was defined by \cite{Hoetal2012} - areas are removed due to a number of reasons, such as image quality and Galactic extinction - using a HEALPix\footnote{HEALPix URL site: {\tt http://healpix.jpl.nasa.gov}} \citep{Gorskietal2005} mask with resolution Nside$=$1024. For the characterization analysis of the combined LOWZ and CMASS samples and the study of their auto-power spectrum, we propagate this mask to Nside$=$128 and then remove all data in HEALpix pixels with weight less that 0.85. This procedure leaves a sample of $1.6\times10^6$ galaxies occupying a $9,255$ deg$^2$ sky footprint. The sample's redshift distribution, determined empirically from the BOSS spectroscopic redshift distribution, is displayed in the left panel of Fig. \ref{fig:redd}.
For the cross-correlation analysis with CMB WMAP9 data, the mask and the CMB and galaxy maps were further downgraded to a resolution parameter Nside$=$64, which corresponds to a pixel area of slightly below one square degree. In this case, in order to avoid mask border effects, only masked pixels with amplitudes above 0.9 were considered. In combination with the WMAP9 sky mask, this cut leaves $1.4\times 10^6$ galaxies on $7,660$ deg$^2$, which represents an effective sky fraction of $0.19$.

\section{WMAP CMB data}
\label{sec:WMAP}
The {\it Wilkinson Microwave Anisotropy Probe} (WMAP\footnote{WMAP URL site: {\tt http://map.gsfc.nasa.gov}}) scanned the microwave sky from 2001 until 2010 in five different
frequencies, ranging from 23 GHz up to 94 GHz. The angular resolution
in each band increases with the band central frequency, but it remains better than
one degree in all bands (for the lowest frequency channel under consideration in our study,
centred at 41\,GHz, the square root of the beam solid angle equals $0.51\degr$). The ISW and its cross-correlation to tracers of the density field arise on larger scales ($\theta > 3\degr$ or $\ell < 80$),  hence angular resolution will not be an issue in this context. 

The WMAP measurements have provided all-sky maps in which, beyond the CMB radiation, other diffuse components and point sources have been identified and characterised, \citep[see, e.g., ][for the latest results from the 9 years of observations]{Bennettetal2012}.
For the data release corresponding to the first 7 years of observation, the signal-to-noise ratio was greater than one for multipoles $\ell<919$ \citep{Jarosiketal2010}. At the large scales of interest for ISW studies, the Galactic and
extragalactic foreground residuals (mostly generated by synchrotron radiation, free-free and dust) remained below the 15 $\mu$K level outside the masked regions \citep{Goldetal2011short}. This analysis has been improved in the recent 9 year data release \citep{Bennettetal2012}, in which an all-sky CMB Internal Linear Combination map, obtained from all bands, has been provided together with its covariance matrix. This analysis constitutes an optimal handling on the uncertainties imprinted on the CMB map by the presence of all those contaminants.

In this context, we shall concentrate our analyses on the foreground cleaned maps
corresponding to the Q (41\,GHz), V (61\,GHz) and W (94\,GHz) bands, after
applying a foreground mask KQ85y9 that excludes $\sim$
15\% of the sky. At the scales of interest, instrumental noise lies
well below cosmic variance and/or foreground residuals, and hence will
not be considered any further. The ISW is a thermal signal whose
signature should not depend upon frequency and hence should remain
constant in all three channels under consideration. All WMAP related
data were downloaded from the {\tt LAMBDA} site\footnote{LAMBDA URL
  site: {\tt http://lambda.gsfc.nasa.gov}}.

\section{Theoretical Predictions}
\label{sec:theorpred}

On scales large enough where linear theory applies, the Poisson
equation relates perturbations in the gravitational potential
$\phi_{\vk}$ at any epoch with those of the matter density contrast
($\delta_{\vk}$) at present via
\begin{equation}
\phi_{\vk} (a) =  - 4\pi G \rho_{b,0} \frac{D(a)\delta_{\vk}}{a\; k^2} ,
\label{eq:poiss1}
\end{equation}
where $a$ denotes the cosmological scale factor, $D(a)$ is the density
linear growth factor, $G$ is Newton's gravitational constant, $\vk$ is
the {\em comoving} wavevector and $\rho_{b,0}$ is the present
background matter density.  From this equation it is easy to see
that for an Einstein-de-Sitter (hereafter EdS) universe, for which
$D(a) = a$, the linear gravitational potentials remain
constant. However, in all other scenarios where $D(a)\neq a$ (suchs as
an open universe, or $\Lambda$CDM, or any other sort of dark energy model, or even a universe with non-negligible radiation energy density)
the potentials will be changing in time. Under these
circumstances it is well known \citep[e.g.,][]{SachsWolfe1967,ReesSciama1968,KofmanStarobinski1985,MartinezGonzalezetal1994} that CMB
photons experience a gravitational blue/redshift, known as the
Integrated Sachs-Wolfe (ISW) effect, i.e.
\begin{equation}
\frac{\delta T}{T_0}(\vnh) = -\frac{2}{c^2} \int_0^{r_{\rm LSS}} dr \;\frac{\partial \phi (r, \vnh)}{\partial r}.
\label{eq:isw1}
\end{equation}
In this equation, $\delta T$ denotes changes in the CMB blackbody brightness temperature,
$\vnh$ refers to a given direction on the sky, and $r$
denotes comoving distance, which relates the time coordinate $t$ and the conformal time $\eta$ via the relation $d \eta = dt/a(t) = \pm dr /c$. The symbol $r_{\rm LSS}$ refers to the comoving distance to the last scattering surface.

\citet{CrittendenTurok1996} first realised that the presence of ISW could be
separated from the CMB temperature anisotropies generated at the
surface of last scattering by cross-correlating the measured CMB
temperature map with density projections of the large scale structure
distribution. The underlying idea is that the gravitational potential wells causing the ISW should be generated by matter overdensities hosting an {\it excess} of galaxies, and hence projected density and ISW fluctuations should be correlated. In the same way, underdense regions in the galaxy distribution ({\it voids}) should also be spatially correlated with gravitational potential hills.
 This correlation can more clearly seen if both the ISW temperature
anisotropies and the galaxy angular number density fluctuations are
expressed in terms of the density contrast Fourier modes
$\delta_{\vk}$. To perform these analysis both fields must be first decomposed on a spherical
harmonic basis, 
\begin{equation}
\delta T^{ISW}(\vnh) / T_0 = \sum_{\ell,m} a_{\ell,m}^{ISW} Y_{\ell,m}(\vnh), 
 \;\;\; \delta n_g(\vnh) / \bar{n}_g = \sum_{\ell,m} a_{\ell,m}^{g}.
 Y_{\ell,m}(\vnh),
\label{eq:d1}
\end{equation}
The multipole coefficients $a_{\ell,m}^{ISW}$, $a_{\ell,m}^{g}$ can then
  be written as \citep[see, e.g.,][]{Cooray2002a,HernandezMonteagudo2008}
  \[
a_{\ell,m}^{ISW} = (-i)^l (4\pi) \int \frac{d\vk}{(2\pi)^3}\;
Y_{\ell,m}^{\star}(\hat{\vk}) \;\times
\]
\begin{equation}
\phantom{xxxxxxxx} \int dr\; j_{\ell} (kr) \frac{-3\Omega_mH_0^2}{k^2} \;\frac{d(D(a)/a)}{dr}   \delta_{\vk},
\label{eq:almISW1}
\end{equation}
\[
a_{\ell,m}^{g} = (-i)^l (4\pi) \int \frac{d\vk}{(2\pi)^3}\;
Y_{\ell,m}^{\star}(\hat{\vk}) \;\times
\]
\begin{equation}
\phantom{xxxxxxxxxxxxx} \int dr\; j_{\ell} (kr)\; W_g(r)\; r^2\; n_g(r)b(r,k)\;D(r) \;
\delta_{\vk} \;\; \bigg/  \;\;\bar{n}_g.
\label{eq:dT1b}
\end{equation}
For each multipole $\ell$, $m$ may take ($2\ell+1$) different values,
ranging from $-\ell$ to $\ell$. $n_g(r)$ denotes the average
comoving galaxy number density at the epoch corresponding to $r$;
whereas $\bar{n}_g$ denotes the average {\em angular} galaxy number
density. The instrumental window function for galaxies at epochs given
by $r$ is provided by the function $W_g(r)$. The functions $j_{\ell}(x)$
correspond to spherical Bessel functions of order $\ell$. In Equation 3, we have 
assumed that the galaxy field is a {\em biased}
tracer of the matter density field, and thus the bias factor $b(r,k)$ may
{\it a priori} be a function of distance/time and/or scale $k$.  If
the galaxy and ISW maps are correlated, then the ensemble average of
the product of the same $(\ell,m)$ multipoles should yield a positive
quantity, i.e., $C_{\ell}^{g,ISW} \equiv \langle a_{\ell,m}^g
(a_{\ell,m}^{ISW})^*\rangle >0$,
\[
C_{\ell}^{g, ISW} = \left( \frac{2}{\pi}\right)\int \;k^2dk\;  P_m(k)
  \; \int dr_1\; j_{\ell} (kr_1) \frac{-3\Omega_mH_0^2}{k^2} \;\frac{d(D/a)}{dr_1} \; \times
  \]
\begin{equation}
\phantom{xxxx}  
  \;\int dr_2\; j_{\ell} (kr_2)\; r_2^2\; W_g(r_2)\;n_g(r_2)b(r_2,k)\;D(r_2),
\label{eq:x_iswrho1}
\end{equation}
where $P_m(k)$ denotes the linear matter power spectrum at
present. The auto power spectra are defined in a similar way,
$C^{X}_{\ell}\equiv \langle a_{\ell,m}^X (a_{\ell,m}^X)^* \rangle$, with
$X,Y=$ISW, CMB, $g$, etc, and the $*$ symbol denotes complex
conjugate.  The detection of this correlation between both maps
(and hence the evidence for ISW) is, however, limited by the fact that
the CMB is dominated by the signal generated at the Last Scattering
Surface at $z\simeq 1,050$. In addition, the galaxy
survey may be affected by shot noise, be incomplete or not probe the
redshift range where the ISW is generated. All these aspects must be
taken into account when computing the theoretical signal-to-noise
ratio (S/N) of the cross-correlation analysis. Since the signals
described so far are written in multipole space, the S/N computation
should be done naturally in multipole space as well. For each multipole $\ell$,
one readily finds that
\begin{equation}
\left(\frac{S}{N} \right)^2_{\ell} =
\frac{\left(C_{\ell}^{g,ISW}\right)^2}{\sigma^2_{C_{\ell}^{g,ISW}}}\times
\mathrm{no.modes} = 
 \frac{\left(C_{\ell}^{g,ISW}\right)^2\times (2\ell+1)f_{sky}}{C_{\ell}^{CMB}(C_{\ell}^g +
   1/{\bar n}_g)+\left(C_{\ell}^{g,ISW}\right)^2}.
\label{eq:s2n1}
\end{equation}
In this expression, the numerator contains the ISW -- galaxy
cross-correlation for multipole $\ell$, while the denominator refers to
the variance associated to its measurement. We next describe this
equation in detail.  Provided that our theoretical description
preserves isotropy, there is no $m$ dependence in the power spectrum
multipoles. Furthermore, since there only exists one single universe
to examine, the averages for each multipole $\ell$ are performed by
considering all corresponding $(2\ell+1)$ $m$ modes (i.e., $\langle
... \rangle \equiv 1/(2\ell+1)\sum_{m} ...$). However, if only a fraction
of the sky $f_{sky}$ is the subject of analysis, then the number of
effective modes for a given multipole $\ell$ shrinks by the same
factor, i.e. $(2\ell+1)f_{sky}$. This explains why the variance for {\em one
  single} estimate of $C_{\ell}^{g,ISW}$ is divided in Eq.(\ref{eq:s2n1})
by $(2\ell+1)f_{sky}$. Although this is a somewhat simplistic description of the
impact of the sky mask on the S/N ratio, this approach has proved to capture
the basics of the S/N degradation under incomplete skies \citep[see,
  e.g., ][]{Cabreetal2007}. 
 
The variance associated to a single estimate of
$C_{\ell}^{g,ISW}$ is given by the product of the CMB and galaxy auto power
spectra plus the squared cross angular spectrum.  The CMB angular power spectrum is the addition of two (uncorrelated) components  produced by the Surface of Last Scattering (SLS)
and the ISW, i.e., $C_{\ell}^{CMB} = C_{\ell}^{SLS} + C_{\ell}^{ISW}$.  On the large angular scales involved, the WMAP instrumental noise in CMB measurements can be safely neglected. At the same time,
provided that galaxies are discrete objects, the galaxy auto power
spectrum contains a term accounting for shot noise. This term
should approximately obey Poissonian statistics and should be proportional to the average
galaxy number density. However, since we are investigating relative
fluctuations of the galaxy angular density field (i.e., $\delta
n_g(\vnh) /\bar{n}_g$ and not $\delta n_g(\vnh)$), the shot
noise term in the power spectrum becomes inversely proportional to
${\bar n}_g$.

\begin{figure*}
\centering
\plotancho{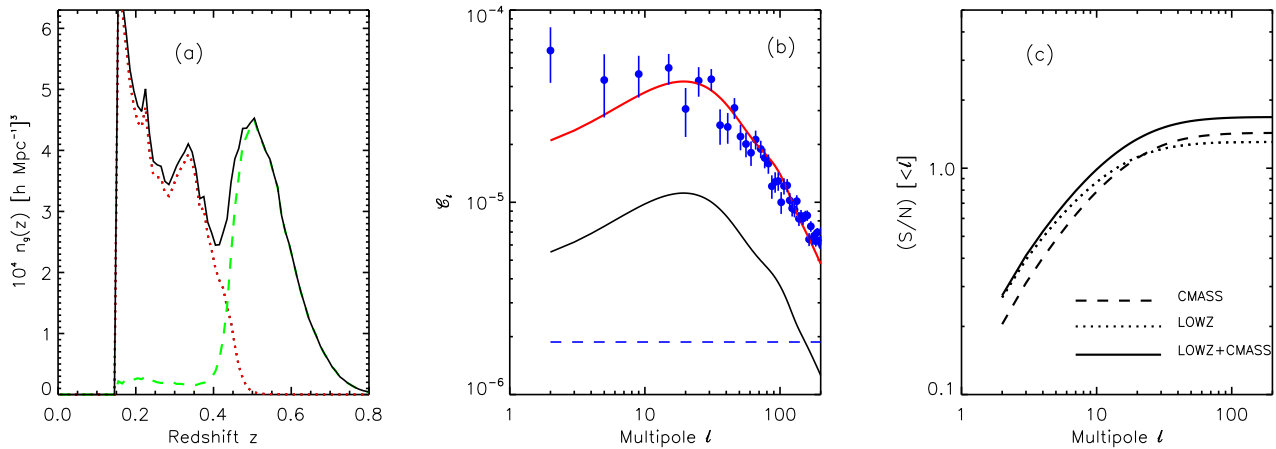}
\caption[fig:redd]{{\it (a)} Comoving LOWZ  (red dotted line) and CMASS (green dashed line)
LG number density as a function of redshift.  {\it (b)} Angular power spectrum estimation of our LOWZ+CMASS sample. The blue dashed line corresponds to the shot noise Poisson term. The solid black line is the linear theory prediction for the $C_{\ell}^g$-s, adopting our fiducial $\Lambda$CDM cosmology model, and assuming the redshift distribution of our LOWZ+CMASS sample (solid line in panel (a)). The red line corresponds to the best fit to the observed angular power spectra (filled blue circles), which yields a constant bias estimate of $b\simeq 1.9$. Error bars are approximated, for each multipole bin, as a fraction $\sqrt{2/(2\ell+1)/f_{\rm sky} / \Delta \ell}$ of the amplitude dictated by the red line, with $\Delta \ell$ being the width of the multipole bin. These bins are centred on their average multipole. See Appendix \ref{sec:aps} for a description of an accurate error bar estimation. {\it (c)}
Cumulative S/N ratio of the ISW -- LG cross power spectrum below a
  given multipole $\ell$.  Dotted, dashed and solid lines correspond to LOWZ, CMASS and LOWZ+CMASS samples, respectively. The latter converges to $\sim1.7$ at $\ell\simeq 60$.}
\label{fig:redd}
\end{figure*}

From Eqs.(\ref{eq:dT1b},\ref{eq:s2n1}) it is clear that the knowledge of
the galaxy redshift distributions and the galaxy bias is required in
order to provide an estimate for the ISW S/N ratio. If the bias is
constant in space and time, then it is easy to show that it cancels out in Eq.(\ref{eq:s2n1})
provided the shot noise term is negligible. From the
LOWZ and CMASS LG samples we know the underlying spectroscopic redshift
distribution, which has been obtained from a spectroscopic subsample, and
is shown in Fig.(\ref{fig:redd}a). Our
LG sample is practically insensitive to the onset of dark energy at
$z \gtrsim1$, but it is able to trace its impact on the growth of potentials at $z<1$, and this redshift range becomes critical when evaluating the ISW contribution. \citet{HernandezMonteagudo2008} found that the maximum contribution to the ISW -- galaxy cross-correlation was produced from 
redshifts in the range $z\in [0.2,1.0]$ when computing the ISW signal in redshift shells of different widths. The comoving number density of the LOWZ sample increases rapidly at low redshift, as displayed in Fig.(\ref{fig:redd}a).  We consider only LGs placed at redshifts above a minimum photometric redshift $z_{\rm min}=0.15$, since our computations following Eq.~\ref{eq:s2n1} above  show that, for a source population following the redshift distribution found for our LG samples, the total ISW S/N ratio possesses a local maximum around $z_{\rm min}$. In doing so, we are effectively applying a cut in photometric redshift whose departure from the spectroscopic one has negligible impact on the S/N estimates. That is, the impact of photometric redshift errors on the estimated S/N lies at the few percent level and hence will be ignored hereafter. Also, below this redshift, the LOWZ sample is known to become contaminated by galaxies from the SDSS main sample, \citep{Parejkoetal2013}

The cumulative ISW S/N ratio is greatly limited by our sky coverage: we must discard low Galactic latitude regions where WMAP CMB data may be contaminated, as well as those regions discarded by the SDSS DR8 survey footprint. Furthermore, since the cross-correlation analyses involving WMAP data require relatively low resolution maps with Nside$=$64, it becomes necessary to remove pixels which lie close to the borders as they drastically affect the estimated galaxy density. When combining both WMAP9's Kq85y9 mask and SDSS LOWZ+CMASS's effective mask, and after dropping all mask pixels with mask value below 0.90, we have an effective sky fraction under analysis of $f_{\rm sky} \simeq 0.19$. Finally, the linear constant bias for our LG sample is computed via a $\chi^2$ fit from the ratio of the measured LG angular power spectrum and the linear theory prediction for the dark matter angular power spectrum, yielding a value of $b\simeq 1.88 \pm 0.11$, see Fig.\ref{fig:redd}(b).

We provide, in Appendix \ref{sec:aps}, a complete characterization of the angular power spectrum of the different galaxy samples, and analyze systematic residuals, study the consistency with the linear Gaussian prediction and obtain constrains on local-type of non-Gaussianity. All these inputs allow the computation of the amplitude of the ISW -- LG cross-correlation that we expect in the concordance $\Lambda$CDM model, and from it the differential (S/N)$^2$ per
multipole $\ell$ as given by Eq.(\ref{eq:s2n1}). In Fig.(\ref{fig:redd}c)
we show the integrated S/N ratio below a given multipole $\ell$,
\begin{equation}
\frac{S}{N} (<\ell) = \sqrt{\sum_{\ell'=2}^{\ell} \left( \frac{S}{N}\right)^2_{\ell'}}.
\label{eq:s2n2}
\end{equation}
The S/N converges at $\sim 1.7$ for $\ell\simeq 60$, as expected from the large scale nature of the ISW signal, \citep{HernandezMonteagudo2008}.
\section{Statistical Methods}
\label{sec:statmethods}

In this section we outline the three different statistical methods we use in our cross-correlation analyses. One is based in real space, another in multipole space, and the third one in
wavelet space. We thus search for the consistency among the outputs of these different methods after blindly testing the null hypothesis. There exist other approaches \citep[like, for instance, the template fitting technique used in ][]{Giannantonioetal2012}, which may yield slightly higher S/N ratios when searching for the ISW under a given model for the galaxy sample.
We end this section by applying these three analyses on mock
ISW -- LG data produced in the concordance $\Lambda$CDM
model. HEALPix\footnote{HEALPix URL site: {\tt http://healpix.jpl.nasa.gov}} \citep{Gorskietal2005} tools were used when implementing these
analyses. In particular, unless otherwise indicated, all maps were
downgraded to $N_{side}=64$ resolution ($\theta_{pixel} \sim 1$\degr).

\subsection{The Angular Cross Power Spectrum (ACPS)}
As mentioned above, this method computes the average product of
multipole coefficients from each (ISW and galaxy) map, i.e.,
\begin{equation}
{\tilde C}_{\ell}^{g,CMB} = \frac{1}{2\ell+1} \sum_m \tilde{a}_{\ell,m}^g (\tilde{a}_{\ell,m}^{CMB})^*.
\label{eq:acps2}
\end{equation}
 In the presence of a sky mask, which excludes un-observed or contaminated
 regions, the estimates of the multipoles $\tilde{a}_{\ell,m}^g$,
 $\tilde{a}_{\ell,m}^{ISW}$ become both biased and correlated\footnote{In
   the standard homogeneous and isotropic scenario
   $\langle \delta_{\vk} \delta_{\vq}^*\rangle=(2\pi)^3P_m(k)
   \delta^D(\vk-\vq)$, and as a consequence for $f_{sky}=1$ we have $\langle
   a_{\ell,m}^{X} (a_{\ell',m'}^Y)^*\rangle = C_{\ell}^{X,Y} \delta^K_{l,l'}
   \delta^K_{m,m'}$; where $\delta^K_{i,j}$ is the Kronecker delta
   ($\delta^K_{i,j}=1$ if $i=j$ and $\delta^K_{i,j}=0$
   otherwise), i.e., {\em different} multipoles are uncorrelated.},
 as do the estimates of the $C_{\ell}^{g,ISW}$-s.  The impact of the
 mask can either be corrected for \citep[see][]{Hivonetal2002} or included when
 comparing to theoretical expectations. The computations of the
 covariance matrix for the ACPS becomes necessary when estimating the
 statistical significance of the measured $\tilde{C}_{\ell}^{g,CMB}$-s. In
 order to take into account in a realistic way the effects associated to
 the effective sky mask on the data, we follow a Monte Carlo
 (MC) approach. Ideally, one perform MC mock catalogs
 for both CMB and LG components. In a first stage, however,
 we shall restrict our MC mocks to the CMB component. These MC
 CMB maps are uncorrelated to our LG catalog, and hence the variance
 of our ACPS estimates is given by
 \begin{equation}
 \sigma^2_{\tilde{C}_{\ell}^{g,CMB}} = \frac{C_{\ell}^{LG\;BOSS} C_{\ell}^{CMB}}{\mathrm{no.modes}}.
 \label{eq:sgCx2}
 \end{equation}
 This variance estimate approaches the denominator of
 Eq.(\ref{eq:s2n1}) if $C_{\ell}^{g,CMB}=C_{\ell}^{g,ISW}$, $(C_{\ell}^{g,ISW})^2 \ll
 C_{\ell}^{CMB} C_{\ell}^{g}$ and the values of $C_{\ell}^{g}$ are close to $C_{\ell}^{LG\;BOSS}$. In
 our $\Lambda$CDM cosmological model all these conditions should be fulfilled;
 if this is the case, running MC on the LG component should not
 appreciably change the variance estimate. Our approach can
 also be understood as a {\it null} hypothesis test, i.e., no existing correlation
 between our SDSS LG catalog and WMAP9 data.  In all cases, monopole and dipole ($\ell=0,1$ multipoles) are removed from the area under analysis.
  
 We may or may not average the estimates of $\tilde{C}_{\ell}^{g,T}$ into multipole bins centered
 around a given multipole $\ell_{\rm b}$, but in either case we attempt to add all the S/N
 within the relevant multipole range, which, according to Fig.(\ref{fig:redd}c), is chosen to
 lie in $\ell\in [2,\ell_{\rm max}]$ with $\ell_{\rm max}=60$. In order to minimize the correlation among multipoles, we choose to bin in 32 bins. From each estimate $\tilde{C}_{\ell_{\rm b}}^{g,T}$, a 
 corresponding $rms$ value is computed from the MC simulations ($\sigma_{\tilde{C}_{\ell_{\rm b}}^{g,T}}$);
using this rms value we calculate the following statistic that adds the S/N contribution from all multipoles:
 \begin{equation}
 \tilde{\beta}_{l} \equiv \sum_{\ell_{\rm b}=2}^{l}\frac{\tilde{C}_{\ell_{\rm b}}^{g,T}}{\sigma_{\tilde{C}_{\ell_{\rm b}}^{g,T}}}.
 \label{eq:beta_acps}
 \end{equation}
As just mentioned above, we display results for 32 multipole bins, but very similar results are obtained when varying the number of bins sampling the same multipole range $\ell \in [2,\ell_{\rm max}]$

 From our MC simulations we obtain an accurate estimate for the $rms$ of this statistics $\sigma_{\beta_{\ell}}$ as well, under the null hypothesis (i.e. $\langle \beta_{\ell} \rangle = 0$). The quoted total S/N is built from the estimate of $\tilde{\beta}_{l_{max}}$ as the ratio
 \begin{equation}
 \frac{S}{N} = \frac{\tilde{\beta}_{\ell_{\rm max}}}{\sigma_{\beta_{\ell_{\rm max}}}}.
 \label{eq:s2n_total_acps}
 \end{equation}

 \subsection{The Angular Correlation Function (ACF)}
 
This cross-correlation estimator, defined in real space, is computed
from the subset of unmasked pixels from
 \begin{equation}
 \tilde{w}(\theta ) = \frac{\sum_{i,j\in [\theta,\theta+d\theta]} \delta T(\vnh_i) \delta_g (\vnh_j)}
   	{\sum_{i,j\in [\theta,\theta+d\theta]} 1},
 \label{eq:acf1}
 \end{equation}
 where the sum runs over the subset of pixels $i,j$ that lie at a
 distance $\theta$.  This method is related to the previous one:
 it can easily be shown that the full sky theoretical expectation for
 the ACF can be written in terms of the ACPS as
\begin{equation}
 w(\theta ) = \sum_{\ell} \frac{2\ell+1}{4\pi} C_{\ell}^{g,ISW} P_{\ell}(\cos \theta),
 \label{eq:acf2}
 \end{equation}
 with $P_{\ell}(x)$ the Legendre polynomial of order $\ell$. From this
 expression, it is easy to see that ACF estimates at different angular
 separations ($\theta$-s) will be highly correlated, since they are
 merely the sum of the {\em same} underlying $C_{\ell}^{g,ISW}$-s with
 different weights (provided by the Legendre polynomials). The highest
 S/N is expected at zero lag ($\theta=0$), for which all multipoles
 contribute ($P_{\ell}(1)=1\;$ for all $\ell$).  As for the ACPS, in order to
 estimate the $rms$ of the ACF estimates we first run MC on the CMB
 component as a null hypothesis test. If an evidence for a cross-correlation 
is found, then MC on the galaxy template will be
 implemented. Both CMB and galaxy maps are normalized to have zero
 mean in the area under analysis.

\subsection{The Spherical Mexican Hat Wavelet (SMHW)}

\begin{figure}
\includegraphics[width=8cm]{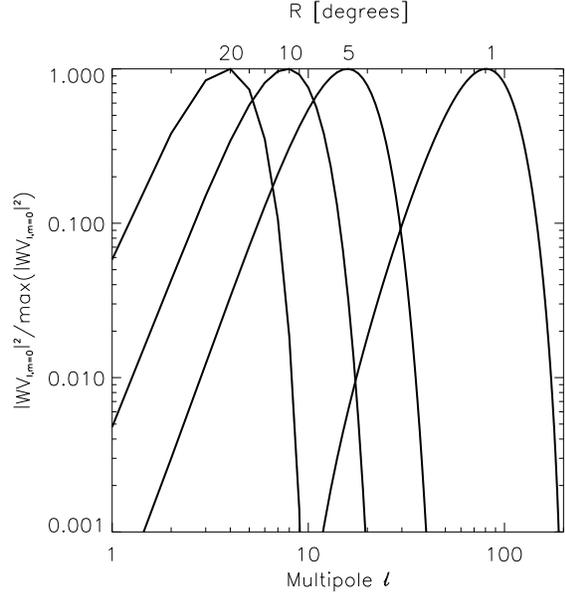}
  \caption{Multipole window function for Spherical Mexican Hat Wavelet of different sizes
    (R=20\degr, 10\degr, 5\degr and 1\degr)}
\label{fig:smhw1}
\end{figure}

The SMHWs form a family of filters defined, in real space, as
 \begin{equation}
wv(R,\vnh) = \frac{1}{\sqrt{2\pi}N(R)}\left[1+\left(\frac{y}{2}\right)^2 \right]^2\left[2- \left(\frac{y}{2}\right)^2\right]\exp{\biggl[-y^2/(2R^2)\biggr]},
 \label{eq:smhw1}
\end{equation}
where $R$ provides the filter scale,
\begin{equation}
N(R) \equiv R\sqrt{1+\frac{R^2}{2}+\frac{R^4}{4}},
\label{eq:smhw1b}
\end{equation}
and $y$ is related to the angular distance $\theta$ by $y\equiv
2\tan\frac{\theta}{2}$. Each scale $R$ defines a limited range of
multipoles to which the filter is sensitive, as shown in
Fig.(\ref{fig:smhw1}). Therefore, by first convolving the temperature
and galaxy maps with these filters and then computing the cross-correlation of 
the resulting maps, one obtains constraints on cross-correlation at
different angular scales. One should, however, keep in
mind that these angular ranges are {\em not} disjoint, and hence
results for different values of $R$ will {\it a priori} be
correlated. For the first implementation of SMHW in ISW studies, see
\citet{Vielvaetal2006}, while further insight in the SMHW properties can be
found in e.g., \citet{MartinezGonzalezetal2002}. This implementation is relatively
simple: multipole coefficients for the wavelet at each scale $R$ are
first computed ($wv_{\ell,m}(R)$). Given the azimuthally symmetric form
of Eq.(\ref{eq:smhw1}), only $m=0$ multipoles contribute and the
convolution is performed by simply multiplying the multipole estimates
$\tilde{a}_{\ell,m}^g$, $\tilde{a}_{\ell,m}^{CMB}$ by the corresponding
wavelet counterpart $wv_{\ell,m=0}(R)$. Maps are inverted back onto real
space, and the zero lag cross-correlation is computed as the average
product outside the mask, by
\begin{equation}
\Pi (R) = \frac{\sum_i \tilde{\delta T}(R,\vnh_i) \tilde{\delta}_g(R,\vnh_i) }{\sum_i 1},
\label{eq:smhw2}
\end{equation}
where $\tilde{s}(R,\vnh)$ denotes the convolved version of map
$s(\vnh)$ with SMHW at scale $R$, and the sum runs only through pixels
{\it outside} the mask\footnote{In practice, the sum is taken only
  through those pixels for which the convolved mask (with the corresponding
  wavelet) is above a given threshold. We show results for a threshold set equal to zero, but the
  results change only by a few percent under a threshold in the range $[0,0.5]$. Our threshold choice is motivated by simulations, which show that  higher S/N is recovered when including all partially masked pixels in the analysis.}. As in previous cases, MC CMB mock maps were produced
and processed exactly as WMAP9 data, and the outcome of these tests
provides the null case probability distribution function for the
$\Pi(R)$-s, to which observed data results are to be compared. In particular, those
simulations provide the $rms$ for each $R$ scale ($\sigma_{\Pi(R)}$).

As for the ACPS, we build the statistic
\begin{equation}
\tilde{\beta}_j = \sum_{i=1}^{j}\frac{\tilde{\Pi}(R_i)}{\sigma_{\Pi(R_i)}},
\label{eq:alpha_wv}
\end{equation}
where the indexes $i$ and $j$ run for the different filter scales $R$ considered. The different
terms in the sum of the right hand side of this equation are correlated, although the MC simulations
account for those correlations and are able to provide an estimate of the $rms$ for this statistics $\sigma_{\tilde{\beta}_j}$ . After adding the contribution from all scales, we quote the total S/N ratio obtained with this statistics, i.e.,
\begin{equation}
\frac{S}{N} = \frac{\tilde{\beta}_{\rm nf}}{\sigma_{\tilde{\beta}_{\rm nf}}},
\label{eq:s2n_wv}
\end{equation}
where the subindex ``${\rm nf}$" denotes the total number of filters considered.


\begin{figure*}
\centering
\plotancho{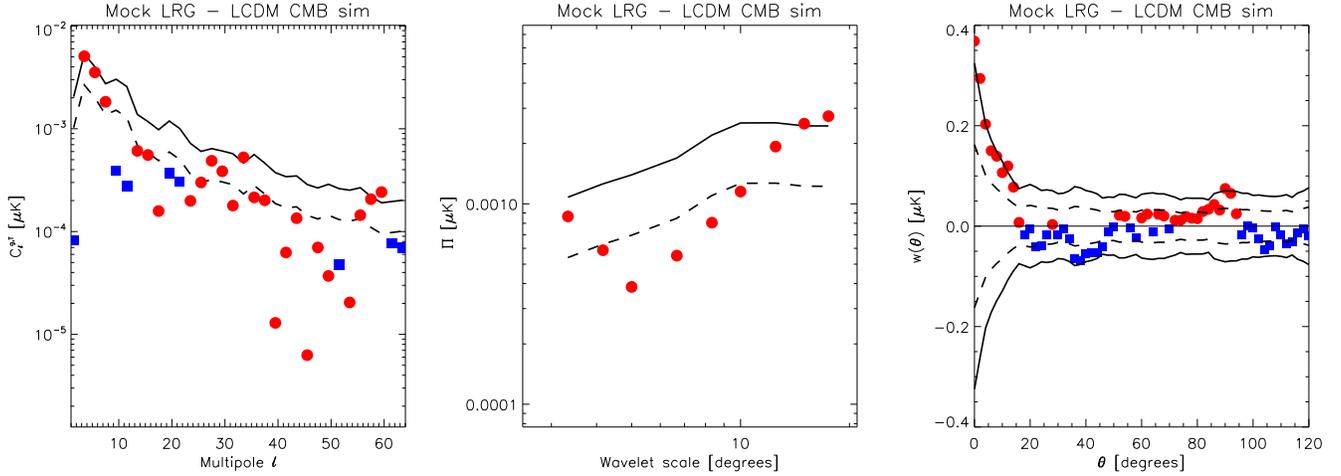}
\caption[fig:ideal1]{Results from a single $\Lambda$CDM simulation; in all panels red (blue) symbols denote positive
  (negative) values. {\it Left panel:} ACPS results for $l\in
  [2,60]$. Solid and dashed lines display the 1, 2\,$\sigma$ confidence
  level after 10,000 MC CMB simulations under the null
  hypothesis. When combining all multipoles, we find an integrated significance at 
  the $S/N\simeq 2.19$. {\it Middle panel:}
  Results for the SMHW test after considering scales of $R=$ 3.3\degr,
  4.2\degr, 5.0\degr, 6.6\degr, 8.3\degr, 10.0\degr, 12.5\degr, 15.0\degr and
  17.5\degr. Again, solid and dashed lines display the 1, 2-$\sigma$
  confidence level after 10,000 MC CMB simulations under the null
  hypothesis. When gathering information at all scales, we obtain
  $S/N\simeq 1.65$. {\it Right panel: } Results of the ACF
  analysis. Dashed and solid lines have same meaning as in previous panels, 
  although these are built upon only 100 MC simulations. At the low $\theta$ range the ACF lies around the 2.2\,$\sigma$ significance level, in good agreement with the other two statistical methods.}
\label{fig:ideal1}
\end{figure*}

\section{Results}

In this section we present the results obtained with the three different cross-correlation methods described above, and is divided in two sub-sections. In the first subsection, we build mock galaxy catalogs following the SDSS-III/BOSS LG clustering and redshift distribution properties (according to our reference $\Lambda$CDM cosmology). These LG mock catalogs have shot noise levels identical to those in real data. Each of those galaxy mocks is correlated to a simulated Gaussian ISW map following the angular power spectrum computed in the same  cosmological model, and each simulated ISW map is added to a Gaussian CMB component according to the angular power spectrum of the CMB anisotropies generated at the surface of Last Scattering. Thus, we have two sets of simulated maps, one for galaxies and one for CMB. Real sky masks are applied when feeding these simulated maps in our analysis pipeline. The second subsection presents the results obtained from observed data, under different levels of corrections for systematics, and are to be compared to the results of the previous subsection. The impact of other artifacts associated to the effect of the mask under use is also addressed in this section, while the description of the attempts to correct for potential systematics is provided in Appendix \ref{sec:syst}. As mentioned above, most of the cross-correlation analyses were performed at the more easily manageable map resolution of Nside$=$64. However, some consistency tests were run under Nside$=$128.

\label{sec:results}
\subsection{Results from $\Lambda$CDM simulations}

Fig.~\ref{fig:ideal1} displays what should be expected for our
cross-correlation analysis in a typical concordance $\Lambda$CDM
simulation. One random LG mock catalog was analyzed in conjunction
with a correspondingly simulated (and correlated) CMB map, under the
same conditions of sky coverage as the observed data. The left panel shows the
results for the ACPS analysis: red circles denote positive values for
the ACPS multipoles, while blue squares correspond to negative
values. 
This convention is followed in the other panels. The dashed
and solid lines correspond to the 1 and 2 $\sigma$ levels,
respectively, obtained after running 10,000 MC CMB simulations under
the null hypothesis, i.e., our LG mock catalog is kept fixed and
CMB skies are generated and cross-correlated to it. This is also the procedure
used when interpreting the results from the SMHW and ACF
tests. A $\chi^2$ computation on our ACPS estimates w.r.t the null hypothesis (for which all $C_{\ell}^{g,ISW}=0$) yields $\chi^2=28.1$. This is obtained after grouping all multipoles in the 32 different bins shown in the figure: about 66\,\% of the null simulations yield higher values of $\chi^2$, so this test does not seem to be too sensitive to the ISW-induced correlation. The $\beta$ statistic defined in Eq.(\ref{eq:beta_acps}) yields a S/N of 2.19 (according to Eq.~\ref{eq:s2n_total_acps}), which reflects the presence of at least five positive points at the 2\,$\sigma$ level in the left panel of Fig.~\ref{fig:ideal1}, together with a majority of positive values of the cross-spectrum estimates.

In the middle panel of Fig.~\ref{fig:ideal1} we display the results for the SMHW coefficients. On all scales the coefficients are positive, and, on the largest scales, they cross the 2\,$\sigma$ level.
In this case, for the nine filter scales under consideration we obtain $\chi^2=13.96$ (about $12$\,\% of the simulations provided higher values of $\chi^2$). The S/N obtained from the $\tilde{\beta}_{\rm nf}$ statistics, introduced in Eq.(\ref{eq:s2n_wv}), yields 1.65, and 4.82\,\% of the 10,000 MC null simulations provided higher values of $\tilde{\beta}_{\rm nf}$.  Finally, the ACF analysis is based upon only 100 MC null simulations, which are nevertheless sufficient to define the 1 and 2$\,\sigma$ levels. This was verified by looking at the error bars after running 500 MC simulations for particular cases, and by computing the error bar at zero lag only ($\theta=0$) after running 10,000 MC simulations. In this case, the ACF goes above the $2\,\sigma$ level in the low $\theta$ range (S/N$=$2.26 at $\theta=0^\circ$, see right panel of Fig.~\ref{fig:ideal1}). 

The values obtained for the S/N in the ACPS and SMHW analysis are in good agreement, within $0.5\,\sigma$, with the theoretical prediction of S/N$\simeq 1.67$ quoted in Section \ref{sec:theorpred}.
After running 10,000 additional MC simulations containing correlated CMB maps and LG mocks, we obtain average S/N values of 1.48 and 1.44 for the ACPS and SMHW algorithms, respectively. Likewise, the average zero lag S/N of 100 mock estimates of the ACF amounts to 1.48, which is in excellent agreement with the other two algorithms. These S/N average values are within 14\,\% of the (approximate) theoretical estimate of 1.67 obtained from Eq.~\ref{eq:s2n2}.

\begin{figure*}
\centering
\plotancho{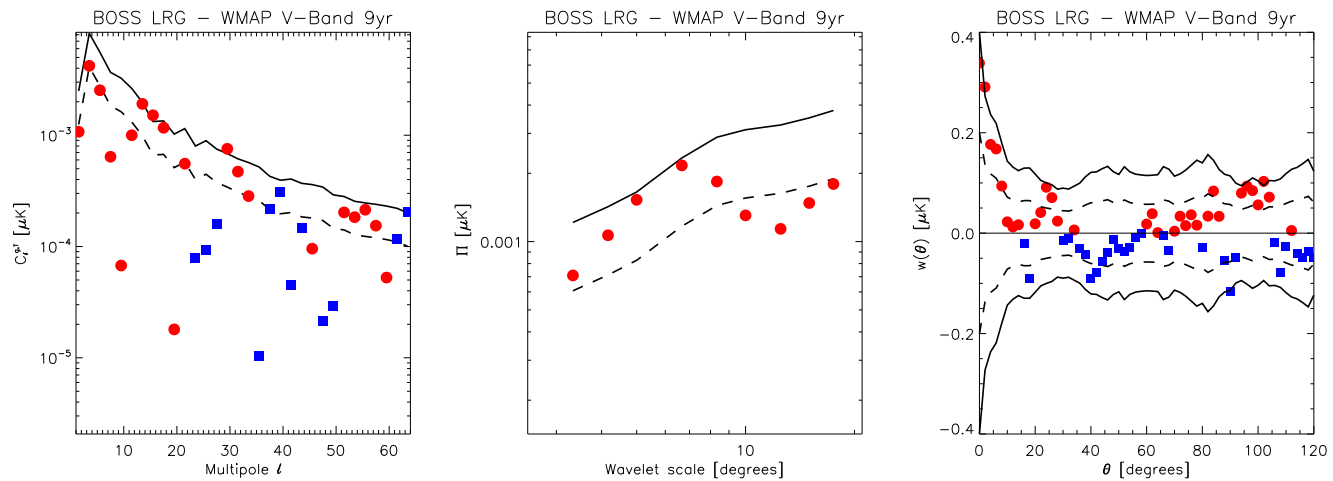}
\caption[fig:real1]{Results from cross-correlating WMAP-9year V-band with LG templates constructed from LOWZ and CMASS samples after correcting for stars, mask value, seeing, sky emission, dust extinction and airmass. In all panels red (blue) symbols denote positive
  (negative) values, and dashed and solid lines display the $1\,\sigma$ and $2\,\sigma$ confidence levels, respectively. Panel organization is identical to Fig.~\ref{fig:ideal1}. In this case, the S/N ratio obtain from the ACPS, SMHW and ACF methods lie at the 1.62 -- 1.67\,$\sigma$ level, in good agreement with the $\Lambda$CDM predictions.}
\label{fig:real1}
\end{figure*}

One further issue to address is to what extent the corrections applied on the LG templates would affect the S/N of our cross-correlation estimators. In Appendix \ref{sec:syst} we describe the procedure by which the LG number density is corrected from possible correlations to known potential systematics (such as stars, sky emission or dust extinction). We use the difference between the raw LG density map and the one accounting for all possible corrections (namely star density, mask pixel value, seeing, sky emission, dust extinction and airmass; see Appendix \ref{sec:syst}). This difference map (Fig.~\ref{fig:syst_corr}) constitutes our template for the systematic impact on our mock galaxy templates. We run again the same set of MC simulations accounting for the correlation between the galaxy mocks and the CMB realisations, but after adding our template for systematics to each simulated galaxy number density map. In this new set of simulations, the average values for the S/N statistics amount to 1.20 and 1.19 for the ACPS and SMWH approaches, respectively; while the average zero lag amplitude of 100 MC mocks of the ACF yield a S/N of 1.03.  This exercise demonstrates that the average impact of the systematics that our analysis has identified and corrected amounts to a 17\,\%--30\,\% decrease of the average S/N, due to a corresponding increase in the measurement errors. Although significant, these changes are not as critical as those induced in the auto power spectra of the galaxy samples at the low $\ell$ regime, (see Figs.~\ref{fig:plotsyst3}, \ref{fig:plotsyst4} in Appendix \ref{sec:syst}).

\begin{figure*}
\centering
\plotancho{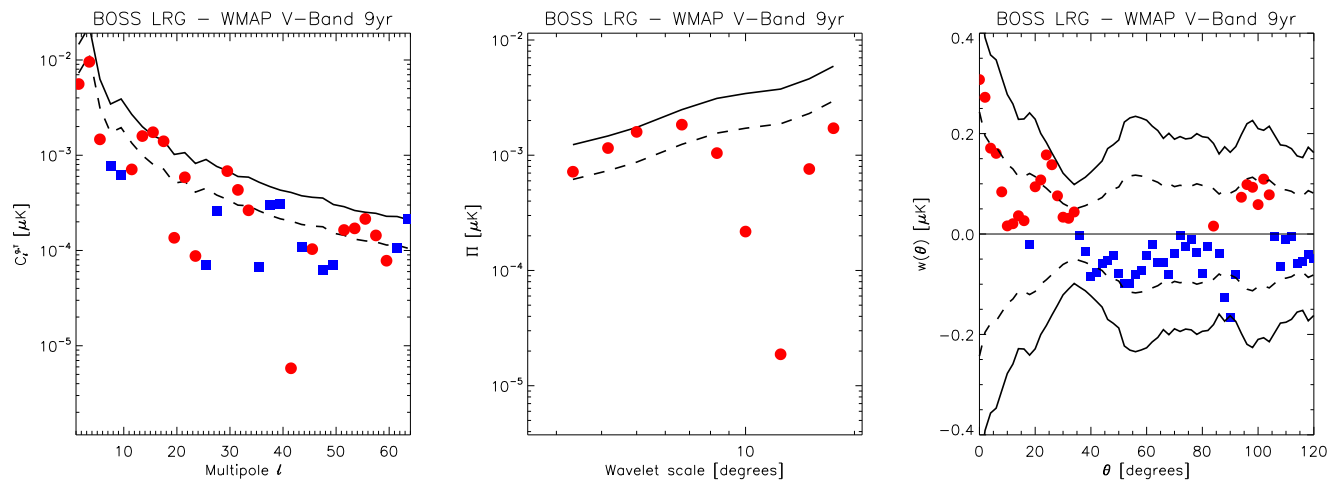}
\caption[fig:real2]{Same as Fig.~\ref{fig:real1}, but after using the {\em raw} LG templates, {\em before} applying any correction. The overall S/N level decreases significantly ($\sim 0.5\,\sigma$) when compared to the corrected one shown in  Fig.~\ref{fig:real1}.   }
\label{fig:real2}
\end{figure*}

\subsection{Results from observations}
\label{sec:resobs}
In agreement with the tests described above, an investigation of the impact of the correction for systematics in the recovery of the S/N of the LG -- ISW cross-correlation shows that, on the observed data set, the corrections for potential systematics described in Appendix \ref{sec:syst} have an effect at the 23\,\% -- 30\,\% level of the total S/N.

In this section, we are showing results for the V band of WMAP, but we have checked that they are very similar to the results obtained in the Q and W bands. In Fig.~\ref{fig:real1} we display the results, under Nside$=$64, for the corrected galaxy template. In this case, the $\chi^2$ test yields 38.1 for 32 d.o.f., i.e., not able to rule out by itself the null hypothesis (more than 21\,\% of the null-case MC simulations provided higher $\chi^2$ values). In terms of the $\tilde{\beta}$ statistic for the ACPS, the S/N becomes 1.62 (only 4.85\,\% of the mock simulations provided larger values of this statistic). For the SMHWs,  $\chi^2=9.34$; this test is again less conclusive than the $\tilde{\beta}_{\rm nf}$ statistics, which yields S/N$= 1.67$. Finally, the zero-lag amplitude of the ACF analysis results in S/N$=1.69$, close to the other two estimates. Had no correction for systematics been applied, the attained S/N levels would have been considerably lower. For instance, for the ACPS and SMHW $\tilde{\beta}$ statistics, the S/N values fall to 1.24 and 1.16, respectively, while the zero-lag amplitude of the ACF yields S/N$=1.27$. A comparison between Fig.~\ref{fig:real1} and Fig.~\ref{fig:real2} shows that the artifacts present in the raw maps only slightly bias the cross-correlation estimates, while widening considerably the allowed area for the 1 and 2\,$\sigma$ confidence levels (i.e., increasing the effective error bars of the cross-correlation analysis).

\begin{figure}
\includegraphics[width=9cm]{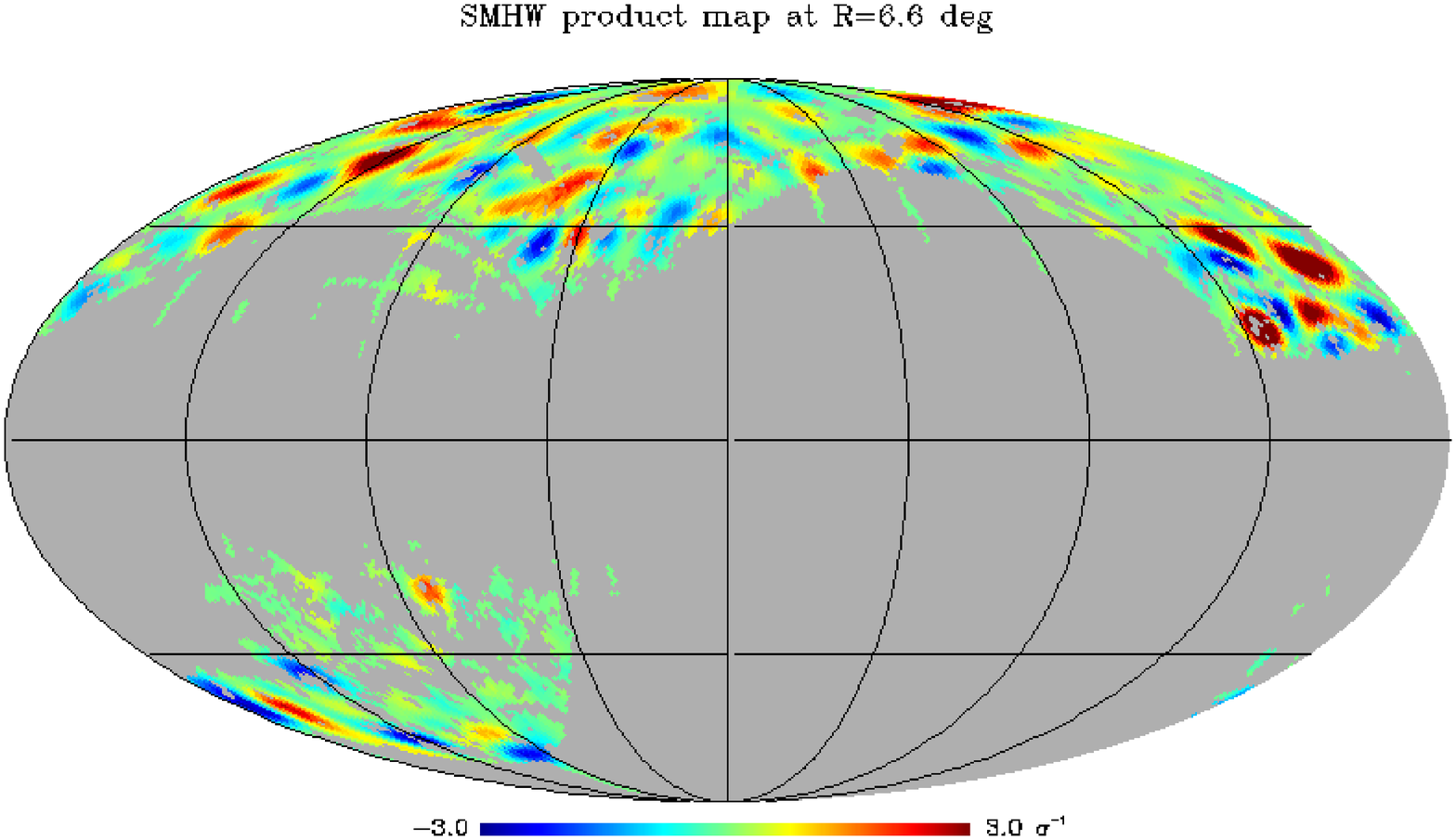}
  \caption{Mollweide projection in Galactic coordinates of the wavelet product map at a scale of R$=6.6\degr$ from WMAP-9yr V and LOWZ+CMASS data. Units are in the $rms$ of this product map, and the central longitude corresponds to $l=0\degr$.}
\label{fig:wvmap}
\end{figure}

The wavelet product maps provide a visual description of the areas where the cross-correlation between the galaxy and CMB maps is built. In Fig.~(\ref{fig:wvmap}) we display the product of the wavelet convolved maps at the scale of 6.6\degr, in units of the $rms$ of the same map. The excess of positive spots over negative ones gives rise to the $\sim 2\,\sigma$ excursion at that scale in the middle panel of Fig.~(\ref{fig:real1}).

We next explore how the recovered S/N levels depend on the mask choice. In our standard mask, we consider for analysis pixels with mask values at Nside$=$64 ranging from 0.9 up to 1. We have checked that making the mask equal to one in all those pixels introduces negligible changes (below the percent level in S/N). Since downgrading the original mask to Nside$=$64 introduces border effects that force a $\sim 16\,$\% decrease of the effective area under analysis, we conduct a single ACPS and SMHW test at Nside$=$128 resolution. At this resolution, the allowed sky fraction where border effects are kept under control increases to $f_{\rm sky}\simeq 0.23$. If S/N truly scales as $\propto \sqrt{f_{\rm sky}}$, then one would expect a $\sim  7$--$10$\,\% improvement of the S/N. In practice, for the corrected LG template we obtain, after 1,000 MC null simulations, S/N values of 1.97 and 1.77 for the ACPS and SMHW tests, respectively. These values constitute an increase of 0.35\,$\sigma$ and 0.10\,$\sigma$ w.r.t the Nside$=$64 case. While the SMHW S/N increase is in perfect agreement with the predicted $f_{\rm sky}$ scaling, the ACPS one is considerably higher ($\sim 0.2\,\sigma$) than expected, although in the appropriate direction. 

The results obtained after implementing an amplitude ($A$) fit to a template, as in \citet{Giannantonioetal2012}, are very similar to the ones presented above. For instance, for the systematic-corrected case addressed in Sect.~\ref{sec:resobs} (and depicted in Fig.~\ref{fig:real1}), the ACPS provides S/N=1.62, while the template fit provides $A/\sigma_A$ = 1.48 (within a 10\,\% difference). Likewise, for the $\Lambda$CDM simulation displayed in Fig.~\ref{fig:ideal1}, the ACPS test provides S/N=2.19, whereas the template fit yields $A/\sigma_A$=2.39, again less than a 10\,\% difference.

Even under Nside$=$128, our results {\em cannot} rule out by themselves an Einstein-de Sitter scenario at the 2\,$\sigma$ level, this is simply a consequence of the signal-to-noise for our samples. Indeed the signal-to-noise we find is in excellent agreement with theoretical expectations based upon the concordance $\Lambda$CDM cosmology (e.g., we find S/N$=1.69$ for the ACF and predict S/N$=1.48$ for the concordance $\Lambda$CDM).

\section{Discussion and conclusions}
\label{sec:disconcl}

After implementing three different statistical approaches to measure the cross-correlation between the CMB and our SDSS-III/BOSS luminous galaxy template, we have found good consistency among their outputs both on simulated and observed data. The mock simulations have shown levels of S/N close to the theoretical prediction of S/N$\simeq 1.67$. In particular, an ensemble of coherent simulations have provided average levels of correlation within 14\,\% of that theoretical estimate, while the analysis of a single random mock realization also yields consistent significance levels within half $\sigma$ from the expected value. The impact that inaccuracies in the systematic removal may have on the obtained S/N is found to be at the $\sim 20$--$30\,\%$ level for our mock galaxy samples, and hence it becomes of relevance when comparing accurate model predictions obtained from numerical simulations with analysis obtained from observational data.

The solid consistency among the three different statistical methods provides evidence for the robustness to the cross-correlation results involving WMAP-9 yr data and SDSS-III/BOSS data. For all three ACPS, SMHW and ACF analyses, the S/N levels are found to lie between 1.62 -- 1.69\,$\sigma$. Furthermore, we have found that not correcting for the impact of systematics on the measured LG number density does change the low $\ell$ angular auto power spectrum estimates; this translates into a lost of S/N at the $\sim 0.5\,\sigma$ level.

Our results on the observational data are in good agreement with the $\Lambda$CDM predictions since they lie within a half $\sigma$ from theoretical expectations, although they do not exclude an Einstein de Sitter scenario by themselves. Our results are in clear tension, however, with results obtained from earlier SDSS data sets. Somewhat peculiarly, the significance of ISW detection in SDSS data sets has tended to decrease as the data set has gotten larger. \citet{Fosalbaetal2003} reported a 3\,$\sigma$ ISW detection using an early release of the SDSS main galaxy sample and \citet{Cabreetal2006} reported 4.7\,$\sigma$ detection using SDSS DR5 data. More recent claims from \citet{Hoetal2008} (at $\sim 2.46$\,$\sigma$ utilizing high redshift SDSS LRGs) or \citet{Giannantonioetal2012} ($\sim 2.5\,\sigma$ adopting the LRG DR7 Sloan sample) are more modest, but still fall $\sim 0.7\,\sigma$ above the theoretical predictions inferred from our photoz SDSS DR8 sample, which is a deeper and wider tracer of the large scale gravitational potentials. Actually, \citet{Giannantonioetal2012} point out that their results with LRGs from SDSS DR7 are about 1.3\,$\sigma$ above the theoretical S/N ratio expected for that survey. On the other hand, the `SDSS' and '2slaq' samples analyzed in \citet{Sawangwitetal2010} are similar in depth to our LOWZ and CMASS samples and their ACF measurements for those two samples agree well with $\Lambda$CDM predictions. In a parallel analysis \citet{Giannantonioetal2013} analyzed a DR8 LRG sample similar to our own, and found a similar significance to our own (1.8) and this result is in good agreement with $\Lambda$CDM predictions.

The reason for decrease in the ISW significance obtained with DR8 data is unclear. While in previous cases it was speculated that the excess anisotropy power on large scales may be responsible for an increased level of evidence for the ISW, in our case we meet the opposite situation, i.e., the presence of artifact systematics increases the level of anisotropy on large scales, and consequently also increases the errors/uncertainties in the recovery of the cross-correlation coefficients in that angular regime.  The amplitude of the cross-correlation is not so critically modified,  and thus we conclude that, at least for our SDSS DR8 data set,  the presence of artifacts in the galaxy template tends to worsen the ISW S/N constraints. 

In this analysis, we use ACPS, SMHW and ACF measurements and find similar results with each technique. One possible drawback of the use of the ACF as unique cross-correlation analysis is that it is sensitive to {\em all} scales present in the map, in particular to the small ones which, while not carrying much ISW information, are the most subject to point source contamination. For example, \citet{HernandezMonteagudo2010} found that, when cross-correlating a radio survey to WMAP data, the ACF provided much higher levels of cross-correlation than the ACPS, and the significance level {\it (i)} was correlated to the flux threshold applied to radio sources and {\it (ii)} restricted to the smallest scales. This example illustrates the utility of implementing simultaneously Fourier- and real-space based algorithms in ISW studies, although it does not imply that point sources must necessarily be biasing all previous analyses. In particular, in our galaxy sample, we find no evidence for significant point source emission in WMAP-9yr sky maps {as suggested by the fact that our ACF significance is almost identical to the ACPS significance}. 

Another possible reason for the mismatch with respect to some of the previous works may be associated to the particular statistic used to infer the ISW significance level and its dependence on the assumed model describing the clustering of the LGs. In our case we are quoting deviations wrt the null hypothesis and our quotes are insensitive to the theoretical model describing the LG angular power spectrum. In other works \citep[e.g.,][]{Giannantonioetal2012}, a fit to a theoretical prediction for the power spectrum/correlation function of the measurements is performed, and different assumptions in such theoretical modeling may give rise to different S/N quotes.

In our study, we make use of all the relevant information available in the SDSS DR8 to not only provide a clean tracer of the potential wells at intermediate redshifts, but also to make precise predictions on what should be seen with this template in an ISW -- cross-correlation study in the $\Lambda$CDM cosmology. The use of Gaussian simulations combined with Poisson sampling should further refine the predictions for the level of ISW signature on the large scale structure. Although our results are in good agreement with expectations, they do not provide conclusive evidence for the presence of ISW.

In practical terms, we conclude that the study of the ISW effect becomes limited by the accuracy to which a given galaxy survey can characterize its source distribution. Uncertainties not only in the galaxy bias, redshift distribution and evolution, but also the impact of sources of systematic uncertainty on large-scales, are currently higher than the uncertainties in the CMB maps, and hence they control the prospects for the ISW characterization. This issue reaches greater relevance when different surveys (each of them with their own uncertainties) are combined in a single ISW cross-correlation study, \citep[e.g.][]{Hoetal2008, Giannantonioetal2012}. The future  of ISW science does not seem to reside so much on the CMB data, but on the side of an accurate cartography of the large scale structure of the universe.

\section*{Acknowledgments}

C.H.-M. thanks L.Verde for useful comments on the draft. C.H.-M. is a {\it Ram\'on y Cajal} Fellow of the Spanish {\it Ministerio de Econom\'ia y Competitividad} and a {\it Marie Curie} Fellow funded via CIG-294183.
F.P. thanks the support from the Spanish MICINNs Consolider-
Ingenio 2010 Programme under grant MultiDark CSD2009-00064
and AYA2010-21231-C02-01 grant. RG-S, JAR-M and CGS acknowledge funding from project
AYA2010-21766-C03-02 of the Spanish Ministry of Science and Innovation
(MICINN). J.X. is supported by the National Youth Thousand Talents Program and the grant No. Y25155E0U1 from IHEP. We acknowledge the use of the Legacy Archive for Microwave Background
Data Analysis (LAMBDA). Support for LAMBDA is provided by the NASA
Office of Space Science. We also acknowledge the use of
the {\tt HEALPix} package \citep{Gorskietal2005}. 
Funding for SDSS-III has been provided by the Alfred P. Sloan
Foundation, the Participating Institutions, the National Science
Foundation, and the U.S. Department of Energy Office of Science. The
SDSS-III web site is {\tt http://www.sdss3.org/}.
SDSS-III is managed by the Astrophysical Research Consortium for the
Participating Institutions of the SDSS-III Collaboration including the
University of Arizona, the Brazilian Participation Group, Brookhaven
National Laboratory, University of Cambridge, Carnegie Mellon
University, University of Florida, the French Participation Group, the
German Participation Group, Harvard University, the Instituto de
Astrofisica de Canarias, the Michigan State/Notre Dame/JINA
Participation Group, Johns Hopkins University, Lawrence Berkeley
National Laboratory, Max Planck Institute for Astrophysics, Max Planck
Institute for Extraterrestrial Physics, New Mexico State University,
New York University, Ohio State University, Pennsylvania State
University, University of Portsmouth, Princeton University, the
Spanish Participation Group, University of Tokyo, University of Utah,
Vanderbilt University, University of Virginia, University of
Washington, and Yale University.

\label{lastpage}

\bibliographystyle{mn2e}
\bibliography{refs}

\appendix
\section{Systematics in the LG data}
\label{sec:syst}

\begin{figure*}
\centering
\plotancho{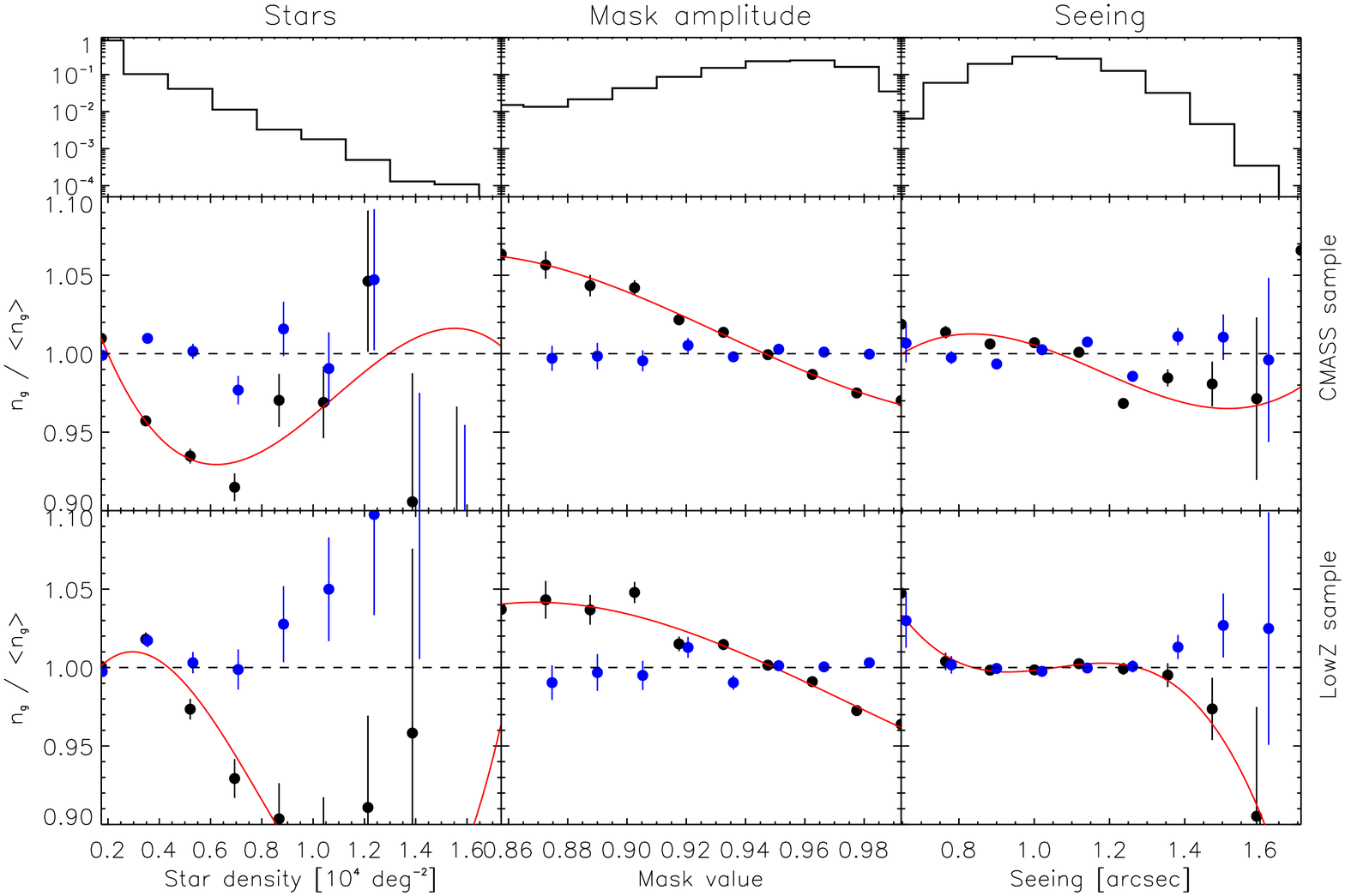}
\caption[fig:plotsyst1]{Scaling of the LG number density versus different potential systematics, namely star density, mask pixel amplitude and seeing (first to third column). Top row panels provide the fraction of the observed area versus the amplitude of the potential systematic, such that the integral below the histograms should equal unity. The middle and bottom rows refer to the CMASS and LOWZ samples, respectively. Black (blue) circles display the scaling before (after) correction (see text). Red solid line displays a spline fit to the measured scaling of the LG number density wrt each potential systematic.}
\label{fig:plotsyst1}
\end{figure*}

\begin{figure*}
\centering
\plotancho{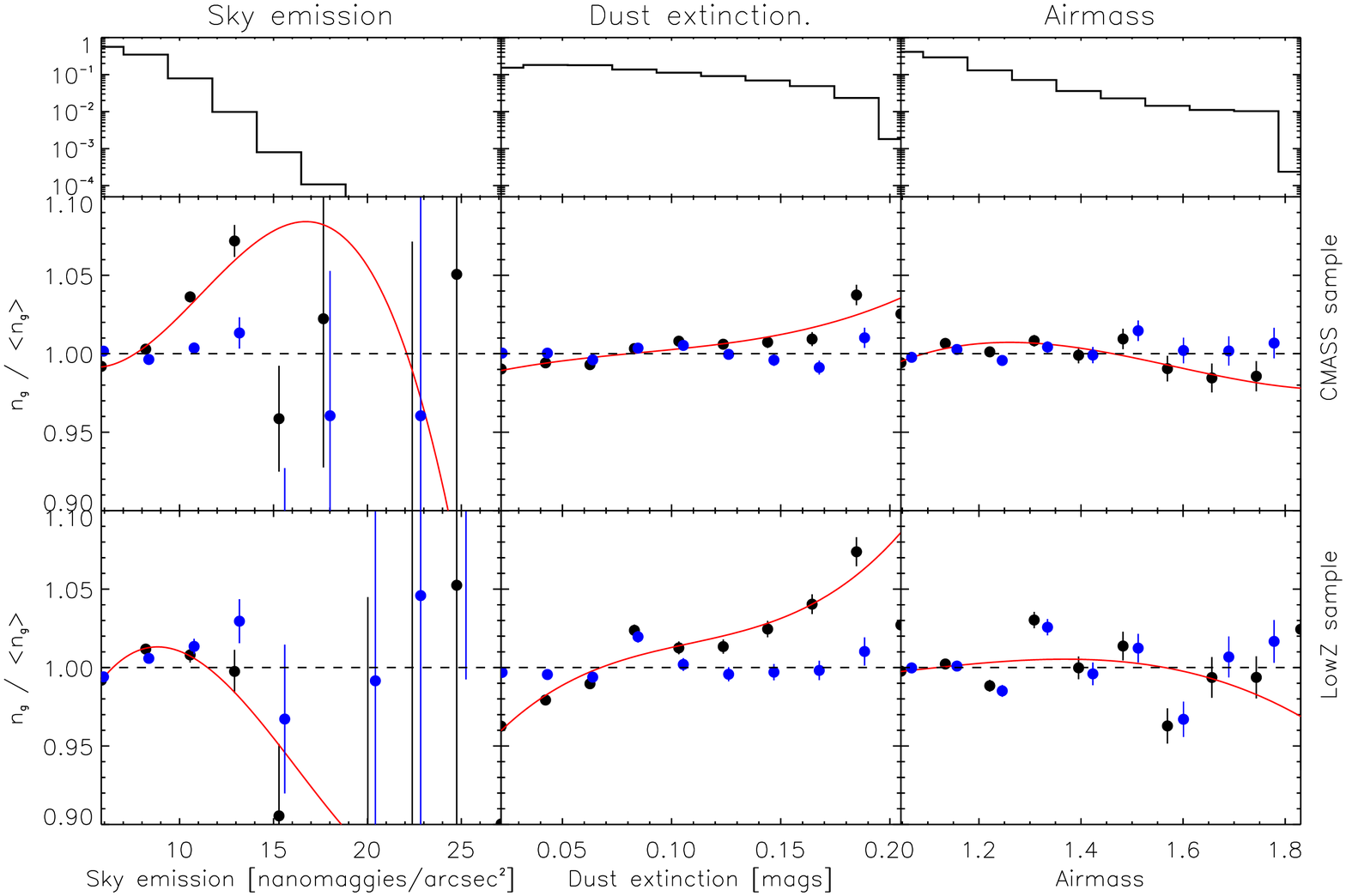}
\caption[fig:plotsyst2]{Same as in Fig.~\ref{fig:plotsyst1}, but referred to sky emission, dust extinction and airmass.}
\label{fig:plotsyst2}
\end{figure*}

\begin{figure*}
\plotancho{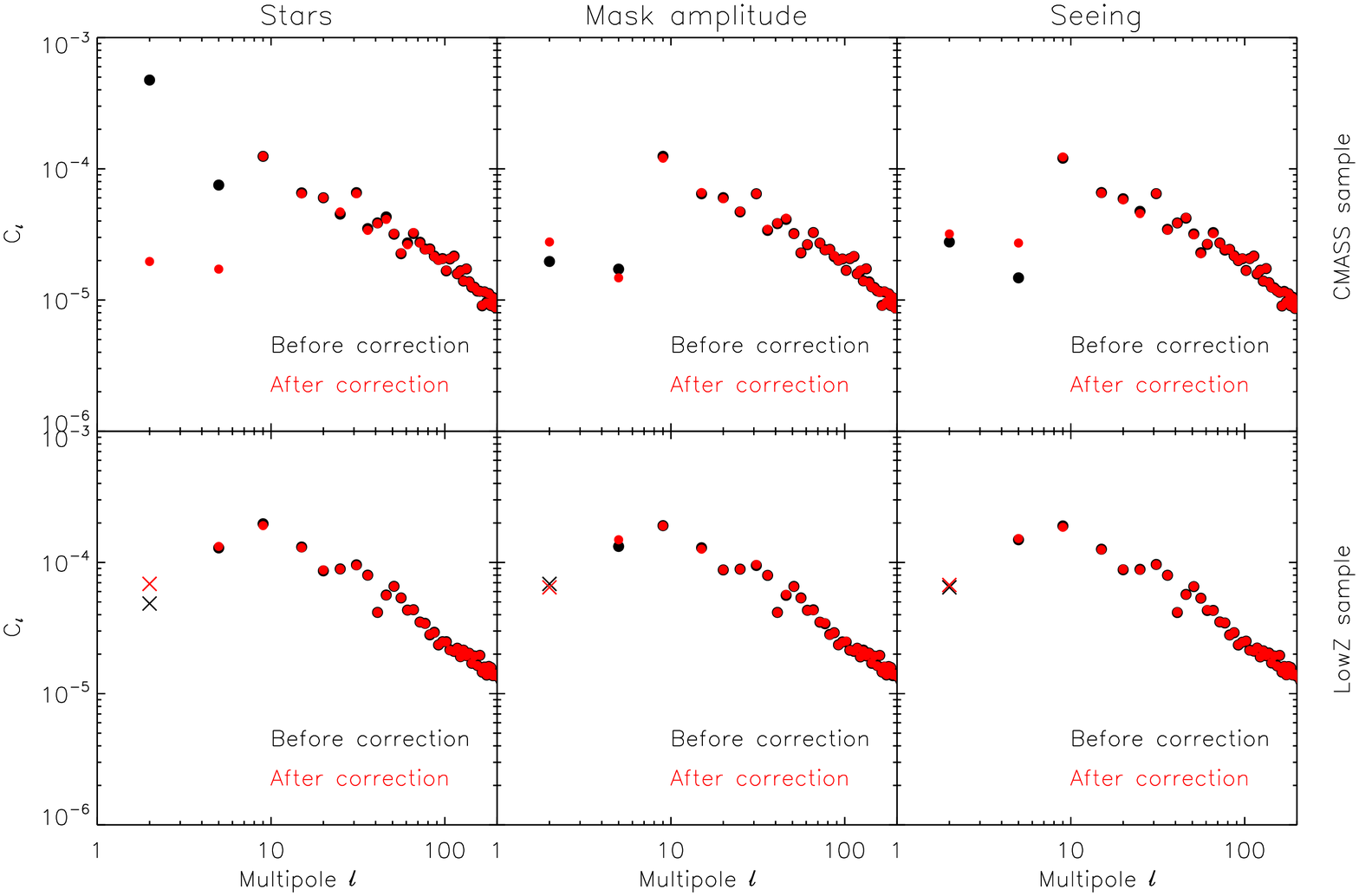}
\caption[fig:plotsyst3]{Impact of the correction for stars, mask pixel value and seeing on the estimated angular power spectra. Black and red circles correspond to angular power spectrum estimates before and after the correction, respectively. Crosses denote {\em negative} values, which may arise after correcting for the mask in the MASTER algorithm.}
\label{fig:plotsyst3}
\end{figure*}

\begin{figure*}
\centering
\plotancho{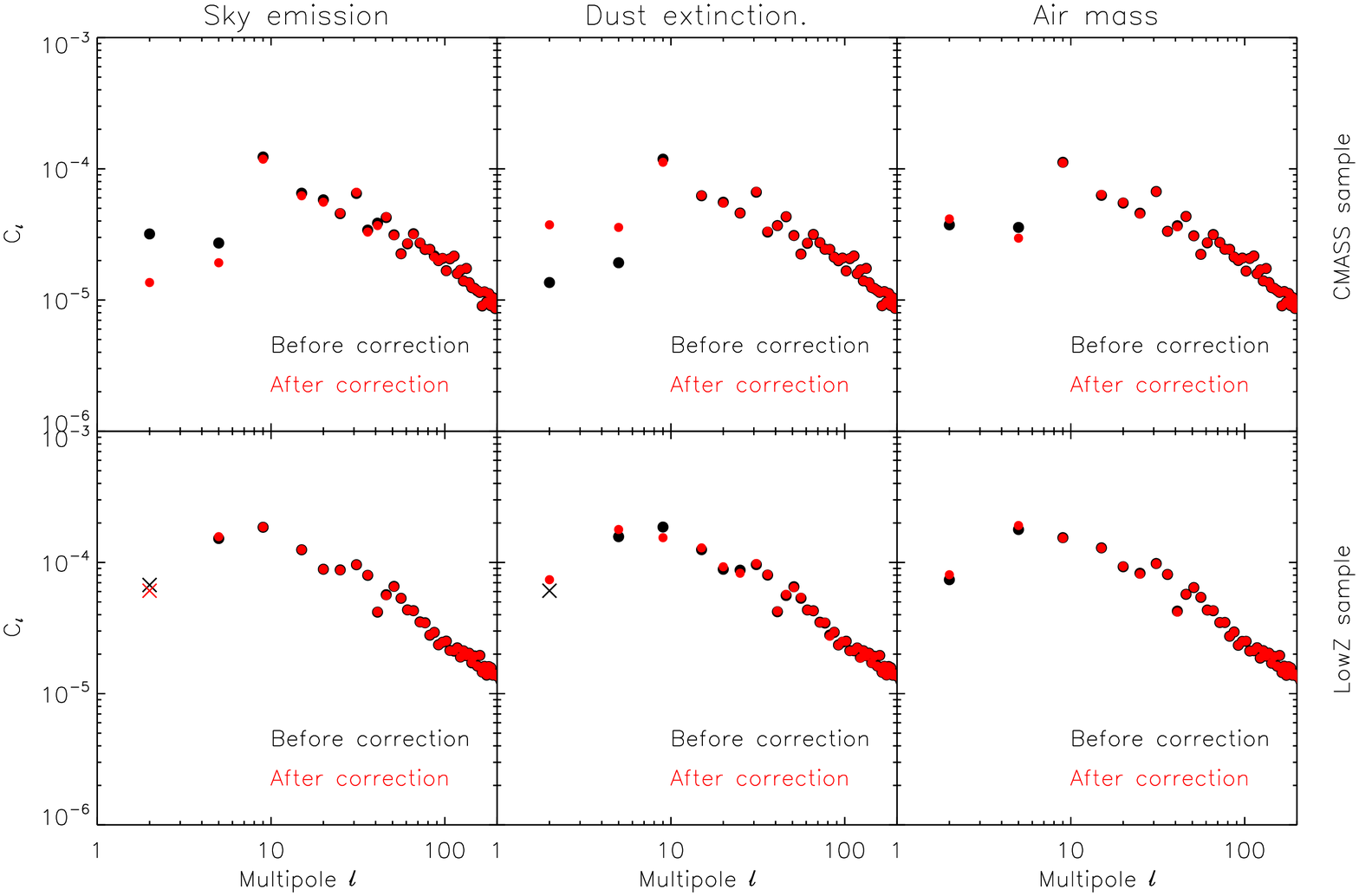}
\caption[fig:plotsyst4]{Same as in Fig.~\ref{fig:plotsyst3} but referred to sky emission, dust extinction and airmass.}
\label{fig:plotsyst4}
\end{figure*}

\begin{figure*}
\centering
\plotancho{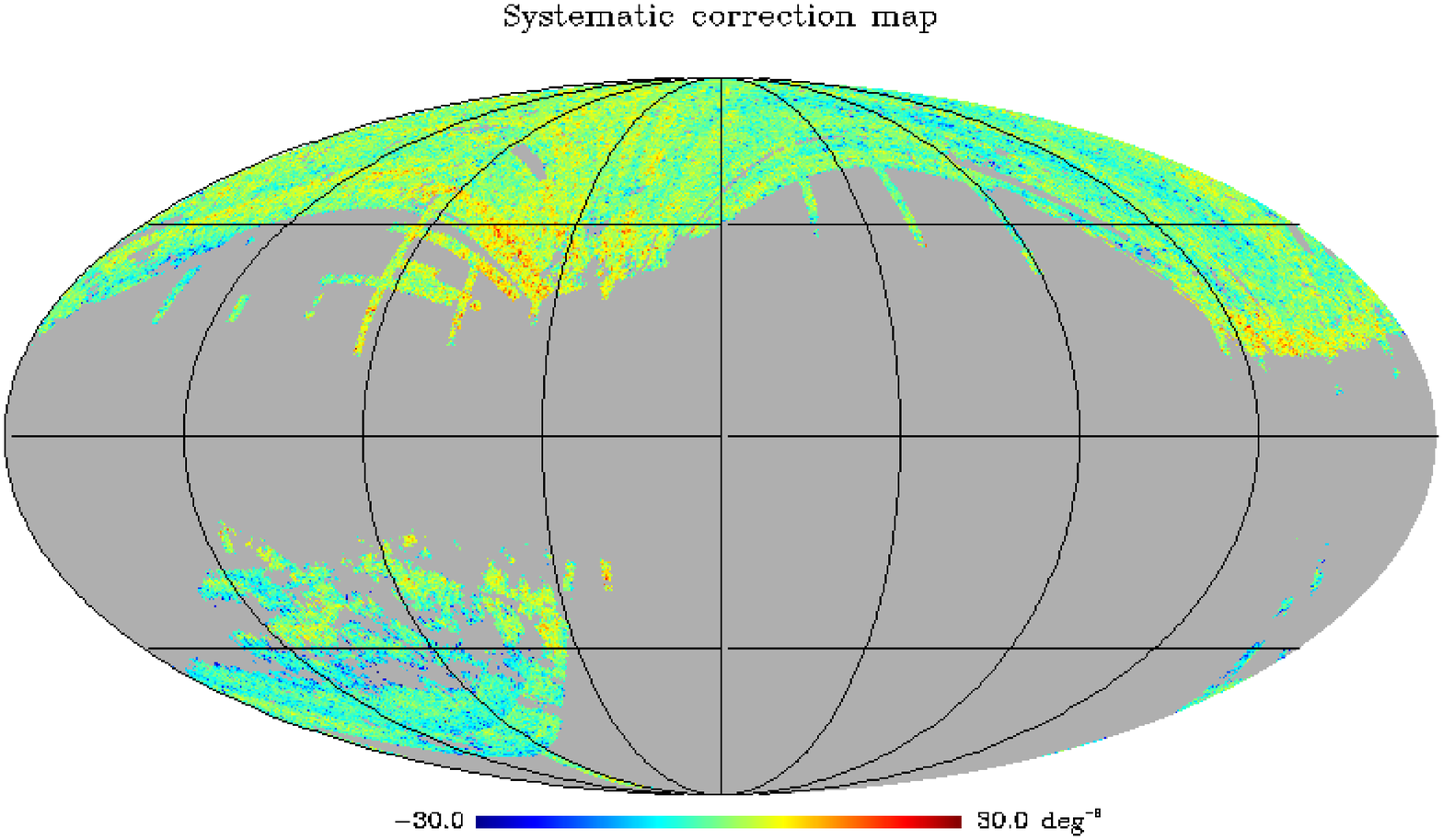}
  \caption{Mollweide projection in Galactic coordinates of the effective correction applied to the combined LOWZ -- CMASS galaxy templates, in units of deg$^{-2}$, with central Galactic longitude at $l=0$. The average galaxy number density in the un-corrected map is close to $160$\,sg.deg$^{-1}$.}
\label{fig:syst_corr}
\end{figure*}

It is well known that the measured SDSS DR8 LG density on the sky is affected by several systematics \citep[see the study for the CMASS sample of ][]{Rossetal2011,Hoetal2012}. The impact of stars is known to be the most relevant, although in this Section we attempt to correct for the biases introduced by other possible systematics such as the effective value of the mask in each pixel, the seeing, the sky emission, the dust extinction and the airmass. The star density template was built upon a star catalogue restricting to the $i_{\rm mod}$ magnitude range $[17.5,\,19.9]$. The mask used in our analysis only considers pixels at $N_{\rm side}=128$ resolution having mask value above 0.85. There exist records of the average seeing at each observed pixel (given in arcsecs), and of the average sky emission in the $i$-band (given in nanomaggies\,arcsec$^{-2}$), together with the dust-induced extinction in the $r$-band, $A_r$, derived from the dust maps of  \citet{Schlegeletal1998}. Finally, we also considered the airmass as potential source of LG number density modulation.

In our approach we investigate the scaling of the LG density w.r.t quantities that are possible sources of systematics, separately for the LOWZ and CMASS samples. When studying the significance of this scaling it is assumed that LGs are Poisson distributed, i.e., we neglect their intrinsic clustering. This approach allows us to assign  error bars to each scaling in a cosmology-independent way, which are however too small, and this must be present when interpreting the scalings.

When building the raw CMASS galaxy density template, we weight each galaxy by its {\it star--galaxy} separation parameter ($p_{\rm sg}$), which provides the probability of a given object to be mistaken by a star, \citep{Rossetal2011}. This parameter was, however, ignored for the LOWZ sample, since in this case there is significantly smaller chance for confusing a galaxy with a star.
In Fig.~\ref{fig:plotsyst1} we study the variation of LG density w.r.t star density, the mask pixel value and the seeing. The top row shows the relative fraction of the observed sky in each bin of each potential systematic (star density, mask value and seeing value), in such a way that the integral under the histogram equals unity. The middle (bottom) row corresponds to the CMASS (LOWZ) sample. In the middle and bottom panel rows, the black filled circles denote the scaling of LG density versus the corresponding potential systematic. The red solid line, in each panel, displays the spline fit to that scaling, and the blue filled circles show the result after correcting the LG density field by the scaling provided by the spline fit. The LG density field corrected for the  ``$X$" potential systematic is given by $n_{\rm g}^{\rm{X_c}} (\vn )$ in
\begin{equation}
n_{\rm g}^{\rm{X_c}} (\vn ) =  \frac{n_{\rm g}(\vn )}{S[X(\vn )]},
\label{eq:systcorr1}
\end{equation}
where $n_{\rm g}(\vn )$ denotes the raw LG number density. The vector $\vn$ denotes a given direction/pixel on the sky.  The symbol $S[X(\vn )]$ denotes the spline fit of the LG number density -- systematic ``$X$" scaling evaluated at the value $X(\vn)$. The normalization for this correction is such that the average LG number density before and after correction are identical. As mentioned above, error bars assume Poissonian statistics. This figure format is identical to that of Fig.~\ref{fig:plotsyst2}, where the correction for sky emission, dust extinction and airmass are considered. 

Each LG density map is corrected for each systematic in succession as displayed from left to right in Figs.~(\ref{fig:plotsyst1},\ref{fig:plotsyst2}). The order at which each potential systematic is corrected should not matter if the effects are independent. However, stars and dust extinction are strongly correlated to Galactic latitude in a similar way, so this analysis requires more caution.  From Figs.~(\ref{fig:plotsyst1},\ref{fig:plotsyst2}) it is clear that stars seem to be significantly modulating the LG number density, since for both LOWZ and CMASS samples the LG number density decreases rapidly with star density (although admittedly the effect is slightly stronger for the LOWZ sample). In either case, the corrected number density seems not to show any remarkable dependence on star density on more than 99\,\% of the observed area (see blue circles below $\sim 900$\,deg$^{-2}$). As shown in the middle panels, the LG number density appears to steadily decrease for increasing values of the mask, suggesting a too conservative estimate of the amount of area masked out around holes present in the footprint. This is common for LOWZ and CMASS samples. When studying this scaling for the degraded Nside$=$64, this scaling is flipped, and at low mask pixel values one finds lower galaxy densities, pointing to a clear footprint border effect. The seeing, conversely, does not appear to significantly modify the LG number density: only for the CMASS sample are there any hints of a trend by which less LGs are identified in regions with worse seeing, although this induces a small correction in the resulting LG number densities.

The left column in Fig.~\ref{fig:plotsyst2} shows that the sky emission modulates the CMASS LG number density in a non-intuitive way: the CMASS number density appears to increase with sky emission. However, this trend is based upon a relatively small fraction of the observed area ($< 8$\,\%), so, if real, it should have a relatively modest impact. The LOWZ sample does not show evidence for such a behavior. On the other hand, dust appears to influence the LOWZ number density but not the CMASS one (see middle column in Fig.~\ref{fig:plotsyst2}). Since dust and stars are spatially correlated on the sky, we explore the scaling of LG density w.r.t dust extinction {\em before and after} correcting for the star density: we find no evidence for any significant change in the LG density versus dust extinction scaling. Only the LOWZ sample scaling suggests that dust may induce confusion in the identification of LOWZ LGs, since the relation of the CMASS number density with dust extinction is essentially flat. Finally, the airmass seems not to introduce a significant trend in the LG number density, as shown by the right column of  Fig.~\ref{fig:plotsyst2}.

In Figs.~(\ref{fig:plotsyst3}, \ref{fig:plotsyst4}) we display the induced changes in the resulting angular power spectra. These spectra are computed after correcting for the bias induced by the sky mask by means of the MASTER approach \citep{Hivonetal2002}. Since we shall perform cross-correlation studies with CMB maps, and in these maps the dipole has no cosmological information, we choose to remove the residual dipole in the effective area in both CMB and LG sky maps using the {\tt remove\_dipole} routine of the HEALPix distribution. In those plots, the correction for stars introduces the largest changes in the estimated angular power spectra, since the quadrupole ($\ell=2$) and the multipole band centred at $\ell=5$ drop by factors of $\sim$ 20 and 6, respectively, for the CMASS sample (changes for the LowZ sample are much more modest). Practically all corrections leave all band power above $\ell=5$ effectively unchanged, while the first two band power spectra display a more unstable behavior. After star correcting the CMASS sample, the
band power centred at multipoles $\ell=2,5$ typically shift between $4\times 10^{-6}$ and $3\times 10^{-5}$. For the LOWZ sample the band power centred at $\ell=5$ barely shifts from $\sim 1.2\times 10^{-5}$, while the one centred on the quadrupole ($\ell=2$) is negative until the dust extinction correction is applied (Negative cases are denoted by crosses rather than filled circles in those figures). The MASTER algorithm may provide negative outputs of auto-power band spectra for cases where these are very poorly constrained (as is the case here). Corrections are applied sequentially, so that the last correction is made upon a map previously corrected for all previous possible systematics.

The total changes in our template for angular galaxy density can be seen in Fig.~(\ref{fig:syst_corr}). Clearly visible is a low Galactic latitude increase in the galaxy density (a compensation for the star-induced galaxy obscuration). Also visible are stripes corresponding to the survey scanning strategy. The average galaxy number density in the corrected map lies close to 160\,sq.deg$^{-1}$.

\begin{figure*}
\centering
\plotancho{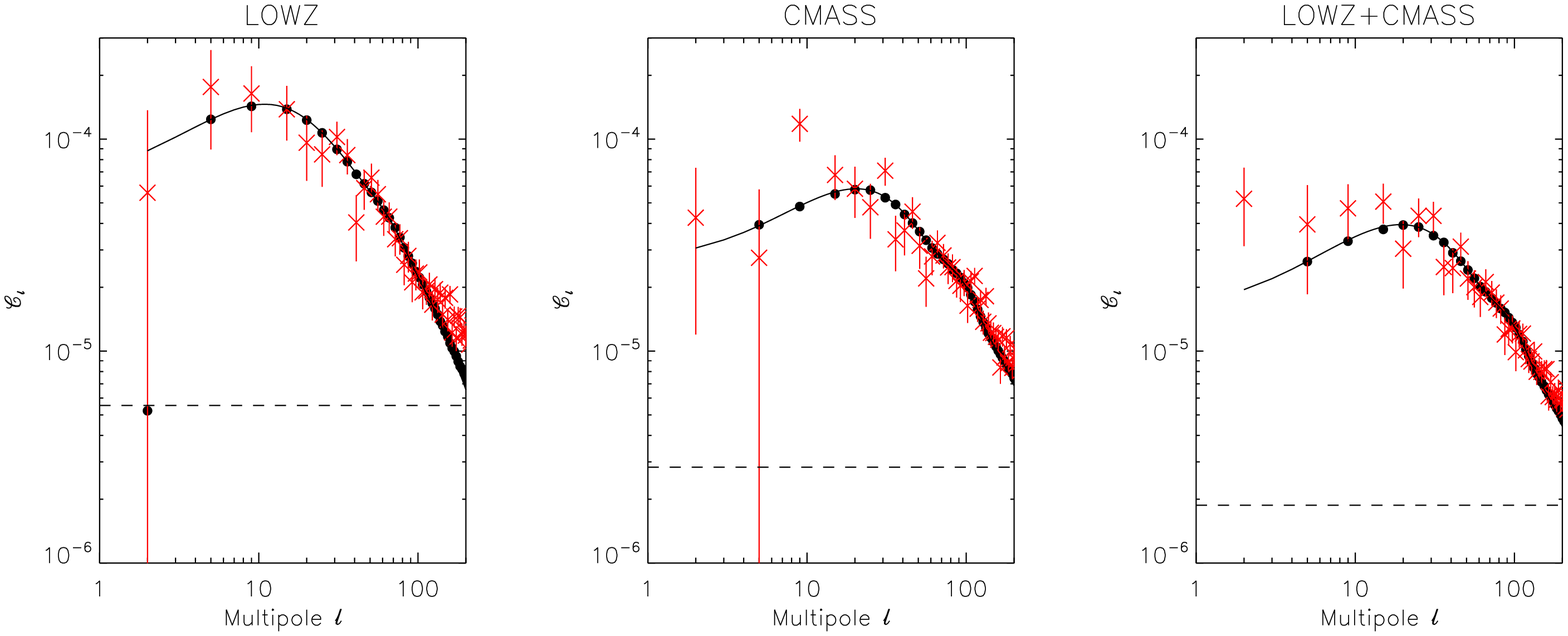}
  \caption{Angular power spectrum estimates for the LOWZ (left panel), CMASS (central panel) and  LOWZ+CMASS (right panel) samples. Observed data are displayed by red crosses, while the average of the output of 1,000 MC simulations are provided by filled black circles. The best-fit linearly biased $\Lambda$CDM prediction is provided by the black solid line; it constitutes the input for the MC simulations. Finally, the shot noise contribution is indicated by horizontal solid lines in each case.}
\label{fig:aps}
\end{figure*}

\section{The angular power spectra of the LG density templates}
\label{sec:aps}
In this Section we address the computation of the angular power spectrum of the LOWZ and CMASS LG samples after conducting the corrections for systematics. We use the MASTER algorithm to correct for the bias induced by the joint sky mask (respecting only a $\sim 23$\,\% of the sky) on the measured angular power spectrum. The outputs of the MASTER algorithm in each of our galaxy samples are displayed by red crosses in the three panels of Fig.~(\ref{fig:aps}) for LOWZ, CMASS and LOWZ+CMASS samples, from left to right, respectively. The entire multipole range is divided in different bins, in which MASTER provides an estimate of the band power spectrum. 

Using the redshift distribution obtained for each of the galaxy samples displayed in the left panel of Fig.~(\ref{fig:redd}), we compute the linear prediction for the angular power spectrum of the angular density contrast. We next fit a constant bias for each galaxy sample: in order to minimize the impact of non-linear effects, we restrict ourselves to the band power spectra contained in the multipole range $\ell \in [2,100]$. Each band power estimate is initially assigned an error equal to $\hat{C}_{\ell} \sqrt{2/(2\ell+1)/f_{\rm sky}/\Delta \ell}$, where $\hat{C}_{\ell}$ is the MASTER output for the band power estimate, $\ell$ is the bin central multipole and $\Delta \ell$ its width. With these error estimates, we perform a $\chi^2$ fit to a constant bias relating the observed power spectra and linear theory predictions in our concordance WMAP-9yr $\Lambda$CDM model. For the LOWZ, CMASS and LOWZ+CMASS samples, we obtain constant bias values of $b=1.98\pm0.11$, $2.08\pm 0.14$ and $1.88\pm 0.11$, respectively. As predicted by the scaling of the bias versus comoving number density of \citet{Nuzaetal2012}, the measured bias values decrease for the sample having higher number density. Although such scaling is referred to the bias computed in space, in our case we find that the joint LOWZ+CMASS sample, having higher angular density than the other two, yields a slightly smaller bias value as computed from the ratio of {\em angular} power spectra. It is nevertheless worth noting that the bias estimates do not deviate from each other more than 2\,$\sigma$. 

With these values for the LG bias we next perform 1,000 MC full-sky Gaussian simulations of  galaxy density, which are re-normalized to have the observed average galaxy number density of each sample. These full-sky Gaussian simulations follow an input angular power spectrum and are produced with the HEALPix tools, which introduce some smearing due to the pixel window function in the simulated maps.  We further run Poissonian realizations of the expected number of galaxies in each map pixel, hence providing our galaxy density templates the corresponding level of shot noise predicted by Poissonian statistics. Each full sky map is multiplied by the effective sky mask used in observed data analysis, and the monopole and dipole are removed from the remaining sky fraction. 

The resulting map is then analyzed with the MASTER algorithm, which produces estimates of the angular power spectrum from our surviving sky fraction. After correcting for the pixel window function and the shot noise term (which equals $1/n$, with $n$ the average galaxy number density in each sample), we obtain the average recovered angular band power spectra estimates displayed by filled black circles in Fig.~(\ref{fig:aps}). These symbols closely follow the linear prediction of the angular power spectra for each sample, after being boosted by the square of the corresponding bias factor (see solid black lines in each panel). Only the first band power estimate (centred at $\ell = 2$) clearly deviates from the linear prediction: this is a consequence of removing the dipole within our effective sky coverage. Since $f_{\rm sky}<1$, the dipole is coupled to immediately higher multipoles, and removing the dipole in our effective sky biases low the first band power estimate. The $rms$ of each band power estimate around the average value is displayed as the effective error bar assigned to the real band power estimates, depicted by red crosses in Fig.~(\ref{fig:aps}). Finally, shot noise levels are depicted, for each sample, by horizontal dashed lines.
 
We next address the question on how consistent is the data with the linearly biased $\Lambda$CDM linear theory prediction for each of the three galaxy surveys. For that, we use the 1,000 MC simulations described above to construct a covariance matrix between the different band power spectra estimates. This allows conducting a $\chi^2$ test for the measured angular band power spectra for each galaxy sample on the (biased) linear theory predictions. When restricting to multipoles below $\ell = 100$, non-linear effects are not important and we obtain, for the 21 band power spectra estimates,  $\chi^2$ values of 13, 30 and 26 for the LOWZ, CMASS and LOWZ+CMASS samples, respectively. For these galaxy samples, we find that 91\,\%, 12\,\% and 26\,\% of the MC simulations yield higher $\chi^2$ values than those produced by the observed data, i.e., the measured angular power spectra are compatible with the linearly biased model predictions. However, when a wider multipole range is considered, then non-linear effects appear and the $\chi^2$ test yields worse results: for all band power spectra below $\ell=154$, 14\,\%, 0.3\,\% and 29\,\% of the MC simulations provide higher $\chi^2$ values. Furthermore, when looking at the entire range $\ell \in [2,205]$, those ratios become 0\,\%, 0\,\% and 3.5\,\% for the LOWZ, CMASS and LOWZ+CMASS samples, respectively. Clearly, by including smaller scales one considerably worsens the fit. Out of the three samples, the most noisy seems to be the CMASS sample, with typically larger $\chi^2$ values, while lower $\chi^2$ values are produced by the LOWZ sample when remaining below $\ell = 100$. 

The observed data points in Fig.~(\ref{fig:aps}) are overall in good agreement with the corresponding linear theory predictions in our Gaussian concordance $\Lambda$CDM model, pointing to a low level of primordial non-Gaussianity. We next restrict our analysis to the combined LOWZ+CMASS sample. We implement the local non-Gaussian bias model used in \citet{Rossetal2013}, which was itself inspired by \citet{Dalaletal2008,MatarreseVerde2008}. In this model, the effective bias is the addition of a (constant) Gaussian bias to a non-Gaussian contribution,
\begin{equation}
b_{\rm eff} = b + (b-1)f_{\rm NL}^{\rm local} \frac{3\delta_c(z)\Omega_{\rm m} H_{0}}{k^2 T(k)c^2},
\label{eq:nGbias}
\end{equation}
where $b$ constitutes the Gaussian bias, $\Omega_{\rm}$ is the total matter critical density parameter, $H_0$ is the Hubble constant and $\delta_c(z) \propto 1/D(z)$ is the spherical collapse overdensity, with $D(z)$ the linear matter density growth factor. The function $T(k)$ is approximated by the dark matter transfer function, $k$ is the wavenumber in units of Mpc$^{-1}$ and $f_{\rm NL}^{\rm local}$ is the local non-Gaussianity parameter. We compute estimates of the angular power spectrum for the LOWZ+CMASS sample for different values of the Gaussian bias  ($b\in[1.70, 2.30]$ and the local non-Gaussianity parameter ($f_{\rm NL}^{\rm local}\in [-300,300]$). We perform a direct computation of the posterior probability for the bias and $f_{\rm NL}^{\rm local}$ after assuming Gaussian priors for the bias and assigning them $rms$ values equal to those found in the $\chi^2$ test described above. We find, however, that our constraints are practically independent of the choice of these priors, and since uncertainties in the amplitude of the anisotropy normalization can also be attached to this parameter, our results are insensitive to this normalization. We restrict our analysis to the multipole range $\ell \in [4,100]$, and ignore the contribution of systematics to the covariance, since even for the band power centred on $\ell = 5$ the extra error induced by uncertainties in the correction procedure is in all but one cases negligible when compared to cosmic variance.  For the covariance matrix of the $C_{\ell}$-s we compute the covariance of the 1,000 MC Gaussian simulations of our $\Lambda$CMD reference model, and scale it to the amplitude of the theoretical angular power spectrum of each model under consideration,
\begin{equation}
\rm{cov}(C_{\ell}^i C_{\ell'}^i) \approx \rm{cov}(C_{\ell}^{\rm ref} C_{\ell'}^{\rm ref}) \times
    \left( \frac{C_{\ell}^i}{C_{\ell}^{\rm ref}} \right) \left( \frac{C_{\ell'}^i}{C_{\ell'}^{\rm ref}} \right),
\label{eq:covmsc}
\end{equation}
where $C_{\ell'}^{\rm ref}$ refers to the theoretical angular power spectrum multipole for our $\Lambda$CDM model, and the superscript $i$ denotes the particular model considered in our two parameter ($b$, $f_{\rm NL}^{\rm local}$) space.
 This expression is equivalent to the assumption that the covariance matrix is dominated by cosmic variance. After taking a flat prior for $f_{\rm NL}^{\rm local}$ in the range $f_{\rm NL}^{\rm local} \in [-300,300]$, we compute marginalized probabilities for both $b$ and $f_{\rm NL}^{\rm local}$. We recover practically the same central value for the Gaussian bias as in the $\chi^2$ analysis outlined above. However, the allowed range shrinks considerably ($b=1.86\pm0.09$ at 95\,\% C.L.). On the other hand, the marginalized probability of the local non-Gaussian parameter peaks on 59 and is contained, at 95\,\% C.L., in the range $f_{\rm NL}^{\rm local} \in [-17,134]$. The width of this allowed range increases by less than 10\,\% when assigning the band power centred on multipole $\ell =5$ a 30\,\% higher uncertainty due to systematics ($f_{\rm NL}^{\rm local} \in [-20,143]$ at 95\,\% C.L.). We next consider including the band power centred on $\ell=2$, despite the concern of it being affected by the dipole subtraction. In this case we allow systematics to increase the measured quadrupole error by 50\,\%, and still the allowed range for $b$ does not change, while $f_{\rm NL}^{\rm local}$ shrinks to $f_{\rm NL}^{\rm local}\in [-2,77]$ at 95\,\% C.L., with the marginalized probability of $f_{\rm NL}^{\rm local}$ peaking on 38. We conclude that the SDSS DR8 LOWZ+CMASS galaxy sample is compatible with the Gaussian scenario. 
 
 The allowed ranges for $f_{\rm NL}^{\rm local}$ quoted here are in general narrower than those quoted in \citet{Rossetal2013} ($f_{\rm NL}^{\rm local}\in [-45,195]$ at 95\,\% C.L.), although in that case the method of handling and marginalizing for systematics is different. Although the works of \citet{Xiaetal2010,Xiaetal2011} may have not considered explicitly the systematics under study in our work, they have excluded very low multipoles from their analysis, and yet they find some weak evidence for positive $f_{\rm NL}^{\rm local}$, \citep[$f_{\rm NL}^{\rm local}=48\pm20$ at 68\,\%C.L.,][]{Xiaetal2011}.  Using a similar set of galaxies, \citet{Giannantonioetal2013} found results consistent with ours, ($-90 < f_{\rm NL}^{\rm local} < 120$ at 95\,\%C.L. for their LRG only sample).

We finally address the issue of the North to South asymmetry. Our mock simulations allow us to construct statistics on the relative difference between the average number of galaxies in the North and South Galactic areas of the survey, for each of the three samples under consideration. On the observed data, we find that the average density mismatch between the North and the South regions amount to -3.7\,\%, -0.8\,\% and -1.9\,\%. \citet{Rossetal2011,Rossetal2012} found similar differences in the North/South number densities, and found that these differences were consistent with the mean North/South photometric offsets determined by \citet{Schlaflyetal2011}. None of the LOWZ mock simulations provide such a high mismatch, while 21\,\% and only 0.2\,\% of the mocks are, for the CMASS and LOWZ+CMASS samples, at similar or higher levels of North-to-South asymmetry. Hence, we conclude that the LOWZ sample shows an excessive number of sources in the south part of the survey whose effects, however, should be largely mitigated in our analysis after removing the dipole in the fraction of area under cosmological analysis.

\end{document}